\documentclass[iop,revtex4,twocolappendix]{emulateapj}

\newcommand{\Kepler}{{\it Kepler}}

\usepackage{color}

\usepackage{makecell}
\usepackage{hhline}
\usepackage{amsmath}

\usepackage{lscape}
\usepackage{amsmath}

\slugcomment{}

\shorttitle{Kepler-538\MakeLowercase{b}}
\shortauthors{Mayo et al.}

\begin{document}


\title{An 11 Earth-Mass, Long-Period Sub-Neptune Orbiting a Sun-like Star}


\author{Andrew W. Mayo\altaffilmark{1,2,3,$\dagger$,$\ddagger$,$\star$}, Vinesh M. Rajpaul\altaffilmark{4}, Lars A. Buchhave\altaffilmark{2,3}, Courtney D. Dressing\altaffilmark{1}, Annelies Mortier\altaffilmark{4,5}, Li Zeng\altaffilmark{6,7}, Charles D. Fortenbach\altaffilmark{8}, Suzanne Aigrain\altaffilmark{9}, Aldo S. Bonomo\altaffilmark{10}, Andrew Collier Cameron\altaffilmark{5}, David Charbonneau\altaffilmark{7}, Adrien Coffinet\altaffilmark{11}, Rosario Cosentino\altaffilmark{12,13}, Mario Damasso\altaffilmark{10}, Xavier Dumusque\altaffilmark{11}, A. F. Martinez Fiorenzano\altaffilmark{12}, Rapha\"{e}lle D. Haywood\altaffilmark{7,$\Box$}, David W. Latham\altaffilmark{7}, Mercedes L\'{o}pez-Morales\altaffilmark{7}, Luca Malavolta\altaffilmark{13}, Giusi Micela\altaffilmark{14}, Emilio Molinari\altaffilmark{15}, Logan Pearce\altaffilmark{16}, Francesco Pepe\altaffilmark{11}, David Phillips\altaffilmark{7}, Giampaolo Piotto\altaffilmark{17,18}, Ennio Poretti\altaffilmark{12,19}, Ken Rice\altaffilmark{20,21}, Alessandro Sozzetti\altaffilmark{10}, Stephane Udry\altaffilmark{11}}

\altaffiltext{$\dagger$}{\texttt{mayo@berkeley.edu}}
\altaffiltext{$\ddagger$}{National Science Foundation Graduate Research Fellow}
\altaffiltext{$\star$}{Fulbright Fellow}
\altaffiltext{$\Box$}{NASA Sagan Fellow}
\altaffiltext{1}{Astronomy Department, University of California, Berkeley, CA 94720, USA}
\altaffiltext{2}{DTU Space, National Space Institute, Technical University of Denmark, Elektrovej 327, DK-2800 Lyngby, Denmark}
\altaffiltext{3}{Centre for Star and Planet Formation, Natural History Museum of Denmark \& Niels Bohr Institute, University of Copenhagen, \O ster Voldgade 5-7, DK-1350 Copenhagen K., Denmark}
\altaffiltext{4}{Astrophysics Group, Cavendish Laboratory, University of Cambridge, J. J. Thomson Avenue, Cambridge CB3 0HE, UK}
\altaffiltext{5}{Centre for Exoplanet Science, SUPA, School of Physics and Astronomy, University of St Andrews, St Andrews, KY16 9SS, UK}
\altaffiltext{6}{Department of Earth \& Planetary Sciences, Harvard University, 20 Oxford Street, Cambridge, MA 02138, USA}
\altaffiltext{7}{Center for Astrophysics ${\rm \mid}$ Harvard {\rm \&} Smithsonian, 60 Garden Street, Cambridge, MA 02138, USA}
\altaffiltext{8}{Department of Physics and Astronomy, San Francisco State University, San Francisco, CA 94132, USA}
\altaffiltext{9}{Department of Physics, Denys Wilkinson Building Keble Road, Oxford, OX1 3RH, UK}
\altaffiltext{10}{INAF - Osservatorio Astrofisico di Torino, via Osservatorio 20, I-10025 Pino Torinese, Italy}
\altaffiltext{11}{Observatoire de Geneve, Universit\`{e} de Gen\'{e}ve, 51 ch. des Maillettes, CH-1290 Sauverny, Switzerland}
\altaffiltext{12}{INAF - Fundaci\'{o}n Galileo Galilei, Rambla Jos\'{e} Ana Fernandez P\'{e}rez 7, E-38712 Bre\~{n}a Baja, Spain}
\altaffiltext{13}{INAF - Osservatorio Astrofisico di Catania, Via S. Sofia 78, I-95123 Catania, Italy}
\altaffiltext{14}{INAF - Osservatorio Astronomico di Palermo, Piazza del Parlamento 1, I-90134 Palermo, Italy}
\altaffiltext{15}{INAF - Osservatorio Astronomico di Cagliari, via della Scienza 5, I-09047, Selargius, Italy}
\altaffiltext{16}{Department of Astronomy, The University of Texas at Austin, Austin, TX 78712, USA}
\altaffiltext{17}{Dipartimento di Fisica e Astronomia ``Galileo Galilei,'' Universita’ di Padova, Vicolo dell’ Osservatorio 3, I-35122 Padova, Italy}
\altaffiltext{18}{INAF - Osservatorio Astronomico di Padova, Vicolo dell’Osservatorio 5, I-35122 Padova, Italy}
\altaffiltext{19}{INAF - Osservatorio Astronomico di Brera, Via E. Bianchi 46, I-23807 Merate, Italy}
\altaffiltext{20}{SUPA, Institute for Astronomy, University of Edinburgh, Royal Observatory, Blackford Hill, Edinburgh EH93HJ, UK}
\altaffiltext{21}{Centre for Exoplanet Science, University of Edinburgh, Edinburgh, EH93FD, UK}

\begin{abstract}
Although several thousands of exoplanets have now been detected and characterized, observational biases have led to a paucity of long-period, low-mass exoplanets with measured masses and a corresponding lag in our understanding of such planets. In this paper we report the mass estimation and characterization of the long-period exoplanet Kepler-538b. This planet orbits a Sun-like star ($V = 11.27$) with $M_*=0.892^{+0.051}_{-0.035} M_\odot$ and $R_*=0.8717^{+0.0064}_{-0.0061} R_\odot$. Kepler-538b is a $2.215^{+0.040}_{-0.034} R_\oplus$ sub-Neptune with a period of $P=81.73778\pm0.00013$ days. It is the only known planet in the system. We collected radial velocity (RV) observations with the High Resolution Echelle Spectrometer (HIRES) on Keck I and High Accuracy Radial velocity Planet Searcher in North hemisphere (HARPS-N) on the Telescopio Nazionale Galileo (TNG). We characterized stellar activity by a Gaussian process with a quasi-periodic kernel applied to our RV and cross-correlation FWHM observations. By simultaneously modeling \Kepler\ photometry, RV, and FWHM observations, we found a semi-amplitude of $K=1.68^{+0.39}_{-0.38}$ m s$^{-1}$ and a planet mass of $M_p=10.6^{+2.5}_{-2.4} M_\oplus$. Kepler-538b is the smallest planet beyond $P=50$ days with an RV mass measurement. The planet likely consists of a significant fraction of ices (dominated by water ice), in addition to rocks/metals, and a small amount of gas. Sophisticated modeling techniques such as those used in this paper, combined with future spectrographs with ultra high-precision and stability will be vital for yielding more mass measurements in this poorly understood exoplanet regime. This in turn will improve our understanding of the relationship between planet composition and insolation flux and how the rocky to gaseous transition depends on planetary equilibrium temperature.
\end{abstract}


\section{Introduction} \label{intro}

To date, nearly four thousand exoplanets have been discovered, but over three quarters of them orbit their host star with periods of less than 50 days (NASA Exoplanet Archive\footnote{\url{https://exoplanetarchive.ipac.caltech.edu/}}; accessed 2019 April 13). However, this is the result of observational biases rather than a feature of the underlying exoplanet population. Bias to short periods is especially strong for the transit method, the most common method of exoplanet detection. Nevertheless, \citet{petiguraetal2018} finds that from 1 to 24 $R_\oplus$, the planet occurrence rate either increases or plateaus as a function of period out to many hundreds of days. Therefore, despite the estimated abundance of long-period planets (i.e., planets with periods longer than 50 days\footnote{We define long-period planets as exoplanets with periods greater than 50 days. This may seem short relative to planets in our own solar system or many of the multi-year period exoplanets already found, but we think it is appropriate, given the relative scarcity of such planets in the known, low-mass planet population.}), our understanding of them is still very incomplete. Relative to the short-period population, there are very few long-period exoplanets (particularly in the low-mass regime) with precise and accurate densities and compositions, and even fewer with atmospheric characterization.

Thus, a larger sample of masses for long-period planets would allow us to address a number of interesting questions. For example, it would allow us to study the rocky to gaseous planet transition and how it depends on stellar flux. We could also investigate planet compositions in or near the habitable zone of Sun-like stars. 

Another interesting feature to study would be the planet radius occurrence gap detected by \citet{fultonetal2017} and \citet{fultonetal2018}. \citet{owenandwu2017} and \citet{vaneylenetal2018} have proposed that photoevaporation strips planets near their host stars down to the core, thus creating the gap. \citet{lopezandrice2018} have investigated the period dependence of the gap position and \citet{zengetal2017} have analyzed the relationship between gap position and stellar type. More long-period planets, with or without planet masses, would provide new insights into the nature and cause of this radius occurrence gap.

In this paper, we characterize the long-period exoplanet Kepler-538b, the only known planet in the Kepler-538 system, first validated by \citet{mortonetal2016}. There is a possible second transiting planet candidate with a period of $117.76$ days, but its existence is very much in question; we briefly discuss this candidate in Section~\ref{cand2}. We determine the properties of the host star, a G-type star slightly smaller than the Sun. We also determine properties of the exoplanet including the orbital period, mass, radius, and density by modeling transit photometry, radial velocity (RV) data, and stellar activity indices. We find that Kepler-538b is the smallest long-period planet to date with both a measured radius and RV mass.

The format of this paper is as follows. In Section~\ref{obs}, we detail our photometric and spectroscopic observations of the planet and its host star. We then discuss stellar parameterization in Section~\ref{stellar_params} and modeling of photometry and spectroscopy in Section~\ref{analysis}. Our results are then presented and discussed in Section~\ref{results_discussion}. Finally, we summarize and conclude our paper in Section~\ref{conclusion}.

\section{Observations} \label{obs}

Photometric observations of the Kepler-538 system were collected with the \Kepler\ spacecraft \citep{boruckietal2008} across 17 quarters beginning in 2009 May and ending in 2013 May. \Kepler\ collected both long-cadence and short-cadence observations of this system. Short-cadence observations (in quarters 3, 7-12 and 17) were collected every $58.89$ s, and long cadence observations (in all other quarters) were collected every $1765.5$ s ($\sim 29.4$ minutes). In particular, we used pre-search data conditioning (PDC) light curves from these quarters downloaded from the Mikulski Archive for Space Telescopes.

Although Kepler-538 was not validated until \citet{mortonetal2016}, it was flagged as a \Kepler\ Object of Interest well before that. As a result, we have conducted a great deal of spectroscopic follow-up on Kepler-538 since it was identified as a candidate host star by the \Kepler\ mission.

First, we collected two spectra with the Tillinghast Reflector Echelle Spectrograph (TRES; \citealt{furesz2008}), an $R=44,000$ spectrograph on the 1.5 m Tillinghast reflector at the Fred Lawrence Whipple Observatory (located on Mt. Hopkins, Arizona). These spectra were collected on the nights of 2010 May 28 and 2010 July 5 and had exposure times of 12 and 15 minutes respectively.

We also downloaded RVs from 26 spectra collected with the HIRES instrument \citep{vogtetal1994} at the Keck I telescope from 2010 July 25 to 2014 July 11. These spectra were originally collected as part of the \Kepler\ Follow-up Observing Program. The standard California Planet Search setup was used \citep{howardetal2010} and the C2 decker was utilized to conduct sky subtraction. Exposure times averaged 1800 s.

Finally, we gathered 83 spectra with the High Accuracy Radial velocity Planet Searcher in North hemisphere (HARPS-N) instrument (\citealt{cosentinoetal2012}, \citealt{cosentinoetal2014}) on the 3.6 m Telescopio Nazionale Galileo (TNG) on La Palma. These observations were made from 2014 June 20 to 2015 November 7, all with exposure times of 30 minutes. They were collected as part of the HARPS-N Collaboration's Guaranteed Time Observations (GTO) program. Using the technique described in \citet{malavoltaetal2017}, we confirmed that none of these spectra suffered from Moon contamination. 

\begin{table*}[t]
\begin{center}
\caption{Stellar parameters of Kepler-538\label{stellar_table}}
\begin{tabular}{llccc}
\tableline
\tableline
Parameter & Unit & SPC & ARES+MOOG & Combined \\
\tableline
\multicolumn{4}{l}{\textit{Stellar parameters}} \\\\

Effective temperature $T_{\mathrm{eff}}$ & K & $5547 \pm 50$ & $5522 \pm 72$ & ... \\
Surface gravity $\log{g}$ & g cm$^{-2}$ & $4.51 \pm 0.10$ & $4.55 \pm 0.12$ & ... \\
Metallicity {[$m/\mathrm{H}$]} & dex & $-0.03 \pm 0.08$ & ... & ... \\
Metallicity {[$\mathrm{Fe}/\mathrm{H}$]} & dex & ... & $-0.15 \pm 0.05$ & ... \\
Radius $R_*$ & $R_\odot$ & $0.8707^{+0.0063}_{-0.0060}$ & $0.8727^{+0.0063}_{-0.0062}$ & $0.8717^{+0.0064}_{-0.0061}$ \\
Mass $M_*$ & $M_\odot$ & $0.925^{+0.034}_{-0.036}$ & $0.870 \pm 0.024$ & $0.892^{+0.051}_{-0.035}$ \\
Density $\rho_*$ & $\rho_\odot$ & $1.404^{+0.061}_{-0.068}$ & $1.31 \pm 0.052$ & $1.349^{+0.089}_{-0.0716}$ \\
Distance & pc & $156.67^{+0.71}_{-0.70}$ & $156.65^{+0.70}_{-0.68}$ & $156.66^{+0.71}_{-0.69}$ \\
Age & Gyr & $3.8^{+2.1}_{-2.0}$ & $6.7^{+1.8}_{-1.6}$ & $5.3^{+2.4}_{-3.0}$ \\
Projected rotational velocity $v\sin i$ & km s$^{-1}$ & $1.1 \pm 0.5$ & ... & ... \\
\tableline
\end{tabular}
\end{center}
\end{table*}

\section{Stellar Characterization}\label{stellar_params}

Stellar atmospheric parameters (effective temperature, metallicity, and surface gravity) were determined in two different ways. First, we combined the two TRES spectra and used the Stellar Parameter Classification tool (SPC; \citealt{buchhaveetal2012}). SPC compares an input spectrum against a library grid of synthetic spectra from \citet{kurucz1992}, interpolating over the library to find the best match as well as uncertainties on the relevant stellar parameters. This method provides a measure for the rotational velocity as well. 

Second, we used ARES+MOOG on the combination of our 83 HARPS-N spectra. More details about this method, based on equivalent widths (EWs), are found in \citet{sousa2014} and references therein. In short, ARESv2 \citep{sousaetal2015} automatically calculates the EWs of a set of neutral and ionised iron lines \citep{sousaetal2011}. These are then used as input in MOOG\footnote{2017 version: \url{http://www.as.utexas.edu/$\sim$chris/moog.html}} \citep{sneden1973}, assuming local thermodynamic equilibrium and using a grid of ATLAS plane-parallel model atmospheres \citep{kurucz1993}. Following \citet{sousaetal2011}, we added systematic errors in quadrature to our errors. The value for surface gravity was corrected for accuracy following \citet{mortieretal2014}. The results from SPC and ARES+MOOG agreed well within uncertainties.

We then estimated stellar mass, radius, and thus density with the \texttt{isochrones} package, a Python routine for inferring model-based stellar properties from known observations \citep{morton2015a}. We supplied the spectroscopic effective temperature, metallicity, the \textit{Gaia} DR2 parallax \citep{gaiacollaboration2016b,gaiacollaboration2018}, and multiple photometric magnitudes (\textit{B, V, J, H, K, W1, W2, W3,} and \textit{G}) as input. Note that we did not use the surface gravity as an input parameter as this parameter is not well determined spectroscopically \citep[e.g. ][]{mortieretal2014}. We ran \texttt{isochrones} four times, using the two different sets of spectroscopic parameters and two sets of isochrones, Modules for Experiments in Stellar Astrophysics (MESA) Isochrones and Stellar Tracks (MIST) and Dartmouth\footnote{The Dartmouth isochrones did not use the \textit{G} magnitude.}.

All four results were consistent, so we followed \citet{malavoltaetal2018} and derived our final set of parameters and uncertainties from the 16th, 50th, and 84th percentile values of the combined posteriors, minimizing systematic biases from using different spectroscopic methods or isochrones. The results of this analysis are listed in Table~\ref{stellar_table}.

As a useful check, we find that our estimates of stellar effective temperature, stellar radius, and distance are all within $1\sigma$ of the \textit{Gaia} DR2 revised \Kepler\ stellar parameters \citep{bergetetal2018}.

\subsection{Consistency with Stellar Activity and Gyrochronology}

As will be discussed in more detail in later sections, RV observations with both HIRES and HARPS-N yielded $\log R'_{HK}$, an indicator of stellar activity. Although $\log R'_{HK}$, like stellar activity, is time variable, taking an average or median over time is still a useful metric of the general activity level of the star. The median $\log R'_{HK}$ with HIRES and HARPS-N was $-4.946 \pm 0.035$ and $-5.001 \pm 0.027$, respectively. The overall $\log R'_{HK}$ across both data sets was $-4.990 \pm 0.034$.

We used this $\log R'_{HK}$ value and the $B-V$ color index\footnote{determined from https://exofop.ipac.caltech.edu/; accessed 2019 July 29} to estimate the stellar rotation period via \citet{noyesetal1984}, finding a value of $32.0 \pm 1.0$ days. Our full model (described in Sections~\ref{phot_analysis} and~\ref{modeling_approach}) included the rotation period as a free parameter, which we estimated to be $25.2^{+6.5}_{-1.2}$ days, in agreement with the stellar activity predicted rotation period to within $1\sigma$. Further, during our processing of photometric data (see Section~\ref{phot_analysis}), we produced a periodogram and an auto-correlation function of the photometry. We found signals near $22$ and $32$ days in the former as well as a weak, broad signal around $20-25$ days in the latter, all of which are near the activity-inferred rotation period or the rotation period estimated from our model.

We also checked that our estimate of stellar age was consistent with gyrochronology. We found a gyrochronological age for Kepler-538 first by determining the convective turnover timescale from \citet{barnesandkim2010} using the $B-V$ color index. Then we used the gyrochronological relation in \citet{barnes2010} to calculate age from the convective turnover timescale and the rotation period (calculated from our full model). In this way, we determined a stellar age of $3.40^{+1.86}_{-0.29}$ Gyr, consistent within $1\sigma$ of our isochrone-derived age of $5.3^{+2.4}_{-3.0}$ Gyr.

\subsection{Possible Binarity of Kepler-538}

In order to investigate whether Kepler-538 may be a binary star or have a companion, either of which could have an effect on the dynamics or nature of Kepler-538b, we downloaded all adaptive optics (AO) and speckle data for the star uploaded to https://exofop.ipac.caltech.edu/k2/ before 2019 July 30. The  Palomar High Angular Resolution Observer (PHARO) on the Palomar-5 m telescope collected AO observations on 2010 July 1 in \textit{J} and \textit{Ks} band; no companions were found between $2$'' and $5$'' down to 19th magnitude. The Differential Speckle Survey Instrument (DSSI) on the WIYN-3.5 m telescope collected speckle observations on 23 October 2010 in \textit{r} and \textit{v} band; no companions were found between $0\rlap{''}.2$ and $1\rlap{''}.8$ down to a contrast of $\Delta m = 3.6$. Finally, the Robo-AO instrument on the Palomar-1.5 m telescope collected an AO observation on 2012 July 28 in the \textit{i} band; no companions were found between $0\rlap{''}.15$ and $2\rlap{''}.5$ down to a contrast of $\Delta m \approx 6$. In short, there is no evidence of a close stellar companion in any of the AO or speckle data.

However, it is worth noting that there is a faint comoving object 17'' from Kepler-538, which \textit{Gaia} found at approximately the same distance of $157$ pc \citep{gaiacollaboration2016b,gaiacollaboration2018}. This means if the two stars are at the same distance, they are separated by $2700 \pm 12$ au, a large enough separation to negligibly affect the planet. Both objects have good astrometric solutions with \textit{Gaia} \citep{lindegrenetal2018}, and their relative motion given by \textit{Gaia} proper motions is $0.408 \pm 0.510$ km s$^{-1}$. However, this relative motion is so slight that we were unable to meaningfully constrain orbital motion.

We estimated the mass of the comoving object to be $0.1169 \pm 0.0075 M_\odot$ by applying the photometric relation in \citet{mannetal2019} to the Two Micron All-Sky Survey (2MASS) \textit{Ks} magnitude \citep{cutrietal2003}, which gives a total system mass of $1.009 \pm 0.044 M_\odot$. With this mass and separation, a circular face-on orbit would have a total relative velocity of $0.576 \pm 0.013$ km s$^{-1}$. Thus, both the velocity of a face-on circular orbit and zero velocity are within $1\sigma$ of the measured relative velocity. With such weak constraints from \textit{Gaia} DR2, we cannot rule out a circular orbit at wide separation nor a highly eccentric orbit, currently observed at apastron, which brings the companion close enough in to potentially affect the planet.

\begin{figure}
\epsscale{1.1}
  \begin{center}
      \leavevmode
\plotone{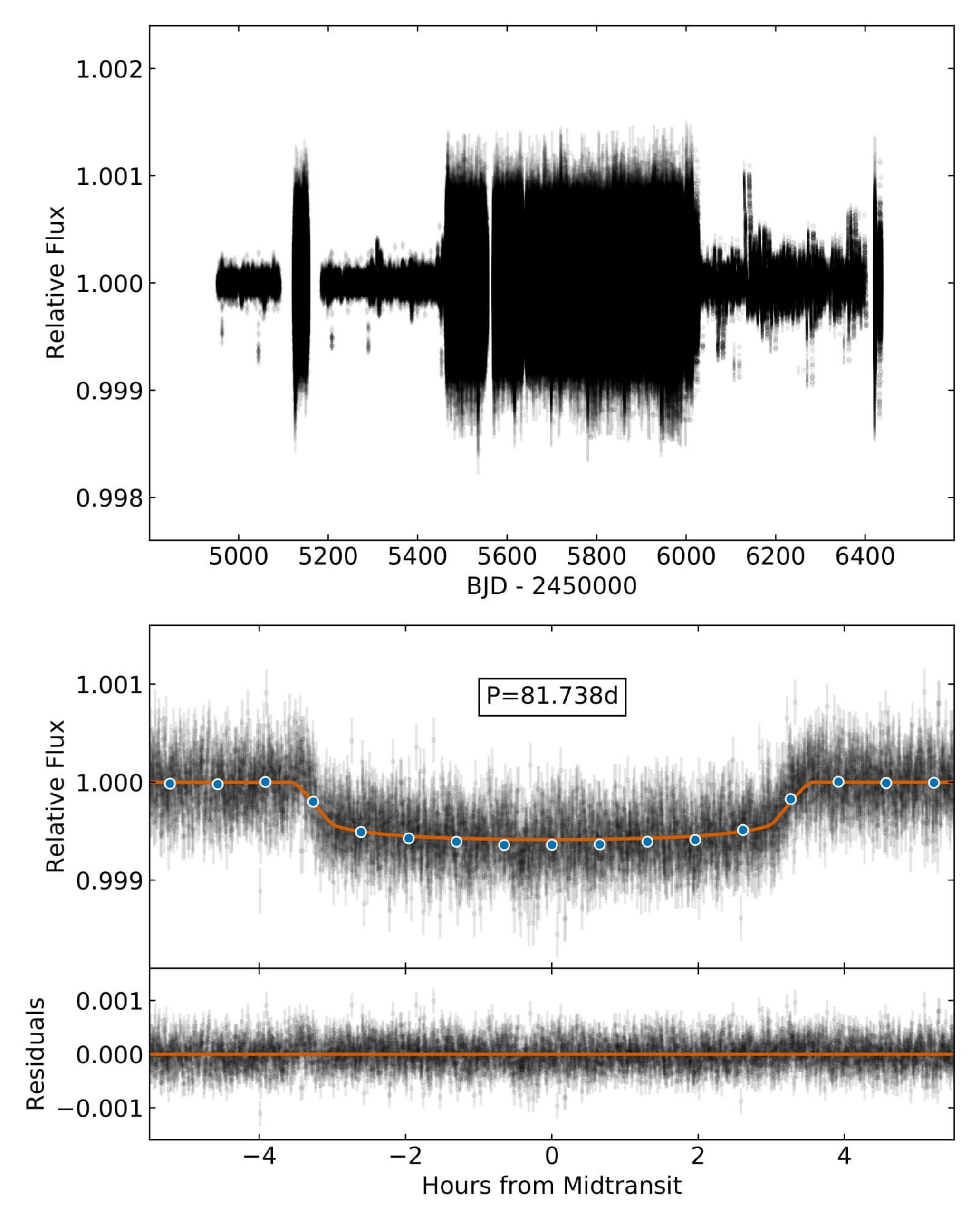}
\caption{Transit plot of Kepler-538b. The top subplot is the pre-search data conditioning (PDC) \Kepler\ photometry. The top panel of the bottom subplot shows the phase-folded photometry in and near the transit of Kepler-538b, with the best-fit transit model in orange and binned data in blue. The bottom panel of the bottom subplot shows the photometric residuals after subtracting the best-fit transit model.} \label{transit_plot}
\end{center}
\end{figure}

\begin{table*}[t]
\begin{center}
\caption{Transit and RV parameters of Kepler-538b \label{results_table}}
\begin{tabular}{llcc}
\tableline
\tableline
Parameter & Unit & This Paper & Priors \\
\tableline
\multicolumn{4}{l}{\textit{Transit parameters}} \\\\

Period $P$ & day & $81.73778 \pm 0.00013$ & Unif($81.73666$,$81.73896$) \\
Time of first transit & BJD-2454833 & $211.6789^{+0.0010}_{-0.0011}$ & Unif($211.6671$,$211.6901$) \\
Orbital eccentricity $e$ & ... & $0.041^{+0.034}_{-0.029}$ ($<0.11$)\footnote{$95\%$ confidence limit} & Beta(0.867,3.03)\footnote{Beta distribution parameter values from \citet{kipping2013a}.}\footnote{Prior also truncated to exclude $e>0.95$.} \\
Longitude of periastron $\omega$ & degree & $140^{+140}_{-90}$ & Unif(0,360) \\
Impact parameter $b$ & ... & $0.41^{+0.10}_{-0.21}$ & Unif(0,1) \\
Transit duration $t_{14}$ & hr & $6.62^{+0.21}_{-0.13}$ & Unif(0,24) \\
Radius ratio $R_\mathrm{p}/R_*$ & ... & $0.02329^{+0.00039}_{-0.00033}$ & Jeffreys(0.001,1) \\
Quadratic limb-darkening parameter $q_1$ & ... & $0.164^{+0.067}_{-0.042}$ & Unif(0,1) \\
Quadratic limb-darkening parameter $q_2$ & ... & $0.74^{+0.16}_{-0.22}$ & Unif(0,1) \\
Normalized baseline offset & ppm & $-2.1^{+2.7}_{-2.8}$ & Unif(-100,100) \\
Photometric white noise amplitude & ppm & $112.2^{+2.5}_{-2.4}$ & ModJeffreys(1,1000,234) \\\\

\tableline
\multicolumn{4}{l}{\textit{RV parameters}} \\\\

Semi-amplitude $K$ & m s$^{-1}$ & $1.69^{+0.39}_{-0.38}$ & ModJeffreys(0.01,10,2.1) \\
HIRES RV white noise amplitude & m s$^{-1}$ & $3.25^{+0.56}_{-0.48}$ & ModJeffreys(0,10,2.1) \\
HARPS-N RV white noise amplitude & m s$^{-1}$ & $2.24^{+0.29}_{-0.27}$ & ModJeffreys(0,10,2.1) \\
HARPS-N FWHM white noise amplitude & m s$^{-1}$ & $6.71^{+0.52}_{-0.46}$ & Jeffreys(0.01,10) \\
HIRES RV offset amplitude & m s$^{-1}$ & $-0.50^{+0.78}_{-0.87}$ & Unif(-5,5) \\
HARPS-N RV offset amplitude & m s$^{-1}$ & $-37322.07^{+0.58}_{-0.73}$ & Unif(-37330,-37315) \\
HARPS-N FWHM offset amplitude & m s$^{-1}$ & $6655.4^{+7.5}_{-8.6}$ & Unif(6600,6700) \\
GP RV convective blueshift amplitude $V_c$ & m s$^{-1}$ & $0.86^{+0.75}_{-0.54}$ & ModJeffreys(0,15,2.1) \\
GP RV rotation modulation amplitude $V_r$ & m s$^{-1}$ & $4.0^{+5.7}_{-3.0}$ & ModJeffreys(0,15,2.1) \\
GP FWHM amplitude $F_c$ & m s$^{-1}$ & $13.3^{+5.9}_{-4.9}$ & Jeffreys(0.01,25) \\
GP stellar rotation period $P_*$ & day & $25.2^{+6.5}_{-1.2}$\footnote{Rotation period uncertainties are highly asymmetric because the posterior includes a large peak at $25$ days and a smaller peak at $31$ days.} & Unif(20,40) \\
GP inverse harmonic complexity $\lambda_p$ & ... & $5.2^{+2.8}_{-2.5}$ & Unif(0.25,10) \\
GP evolution time-scale $\lambda_e$ & day & $370^{+200}_{-140}$ & Jeffreys(1,1000) \\\\

\tableline
\multicolumn{4}{l}{\textit{Derived parameters}} \\\\
Planet radius $R_p$ & $R_\oplus$ & $2.215^{+0.040}_{-0.034}$ & ... \\
System scale $a/R_*$ & ... & $87.5^{+1.5}_{-1.6}$ & ... \\
Planet semi-major axis $a$ & au & $0.3548^{+0.0066}_{-0.0068}$ & ... \\
Orbital inclination $i$ & degree & $89.73^{+0.14}_{-0.06}$ & ... \\
Planet mass $M_p$ & $M_\oplus$ & $10.6^{+2.5}_{-2.4}$ & ... \\
Planet mean density $\rho_p$ & $\rho_\oplus$ & $0.98 \pm 0.23$ & ... \\
Planet mean density $\rho_p$ & g cm$^{-3}$ & $5.4 \pm 1.3$ & ... \\
Planet insolation flux $S_p$ & $S_\oplus$ & $5.19^{+0.31}_{-0.28}$ & ... \\
Planet equilibrium temperature $T_{eq}$ (albedo $= 0.3$) & K & $380$ & ... \\
Planet equilibrium temperature $T_{eq}$ (albedo $= 0.5$) & K & $350$ & ... \\\\

\tableline
\end{tabular}
\end{center}
\end{table*}

\section{Data Analysis} \label{analysis}

Our analysis of photometric and spectroscopic data included a simultaneous fit to both data types. Therefore, we first describe the data reduction process and model components of photometry and spectroscopy separately, then discuss the combined model afterward.

\subsection{Photometric Data} \label{phot_analysis}

We cleaned and reduced the photometric \Kepler\ data using the \texttt{lightkurve} Python package \citep{barentsenetal2019}. Each quarter was cleaned and reduced separately. For a given quarter, observation times without a corresponding flux were removed. Then, a crude light curve model based on the exoplanet parameters reported in the NASA Exoplanet Archive\footnote{\url{https://exoplanetarchive.ipac.caltech.edu/}} (accessed 2019 February 16) was subtracted from the light curve so that in-transit data would not be clipped or flattened out in the next steps. Next, we flattened the light curve using the \texttt{lightkurve} flatten function, which uses a Savitzky-Golay filter. A window length of 615 or 41 was selected (i.e., 615 or 41 consecutive data points) for short-cadence and long-cadence data respectively, which is approximately three times the ratio between the transit duration and the observation cadence. Then, we clipped outlier data points discrepant from the median flux by more than $5\sigma$. Lastly, we added the transit model from the earlier step back to the light curve. The reduced data can be seen in Fig.~\ref{transit_plot}, plotted in time and also phase-folded to the period of Kepler-538b.

We modeled the light curve with the \texttt{BATMAN} Python package \citep{kreidberg2015}, which is based on the \citet{mandelandagol2002} transit model. The model included a baseline offset parameter, a white noise parameter (to allow for instrumental and systematic noise in the data), two quadratic limb-darkening parameters (using the \citealt{kipping2013b} parameterization), the transit time (i.e., reference epoch), orbital period, planet radius relative to stellar radius, transit duration, impact parameter, eccentricity, and longitude of periastron.

We assumed uniform, Jeffreys, or modified Jeffreys priors for most of the parameters in this model, which are listed in Table~\ref{results_table}. A Jeffreys prior is less informative than a uniform prior when the prior range is large and the scale of the parameter is unknown. A modified Jeffreys prior has the following form \citep{gregory2007}:
$$p(X) = \dfrac{1}{X + X_0}\dfrac{1}{\ln({\frac{X_{\mathrm{max}}+X_0}{X_{\mathrm{min}}+X_0}})}$$
where $X_{\mathrm{min}}$ and $X_{\mathrm{max}}$ are the minimum and maximum prior value and $X_0$ is the location of a knee in the prior. A modified Jeffreys prior behaves like a Jeffreys prior above the knee at $X_0$ and behaves likes a uniform prior below the knee; this is useful when the prior includes zero (creating an asymptote for a conventional Jeffreys prior). A Jeffreys prior is simply a modified Jeffreys prior with the knee at $X_0 = 0$.

The only parameter with a different prior was orbital eccentricity. We applied a beta prior to orbital eccentricity using the values recommended by \citet{kipping2013a}; we also truncated the prior to exclude $e>0.95$.

Additionally, we also applied a stellar density prior. This was done given the fact that stellar density can be measured in two distinct ways: from photometry for a transiting exoplanet and from a stellar spectrum combined with stellar evolutionary tracks (we used the latter method in Section~\ref{stellar_params}). Specifically, stellar density can be calculated via the following equation \citep{seagerandmallen-ornelas2003, sozzettietal2007}:

\begin{equation} \label{eqn:density}
    \rho_* = \frac{3\pi}{GP^2}\bigg(\frac{a}{R_*}\bigg)^3
\end{equation}
where the orbital period ($P$) and the normalized semi-major axis ($a/R_*$) are exoplanet properties that can be derived from the light curve. We applied a Gaussian prior to the exoplanet-derived stellar density using the density (and corresponding uncertainties) derived from spectra and stellar evolutionary tracks. 

\subsection{RV Data} \label{rv_analysis}

Our RV analysis of Kepler-538b included not only the RV values determined from our HIRES and HARPS-N spectra, but also a number of indicators of stellar activity estimated from these spectra. For HARPS-N, these included the cross-correlation function (CCF) bisector span inverse slope (hereafter BIS), the CCF full width at half maximum (FWHM), and $\log R'_{HK}$. Our data reduction was performed with the data reduction software (DRS) 3.7 HARPS-N pipeline which applied a G2 stellar type mask. For HIRES, RVs are estimated with an iodine cell rather than cross correlation, so $\log R'_{HK}$ was calculated but not BIS or FWHM.

\begin{figure*}[ht!]
\epsscale{0.8}
  \begin{center}
      \leavevmode
\plotone{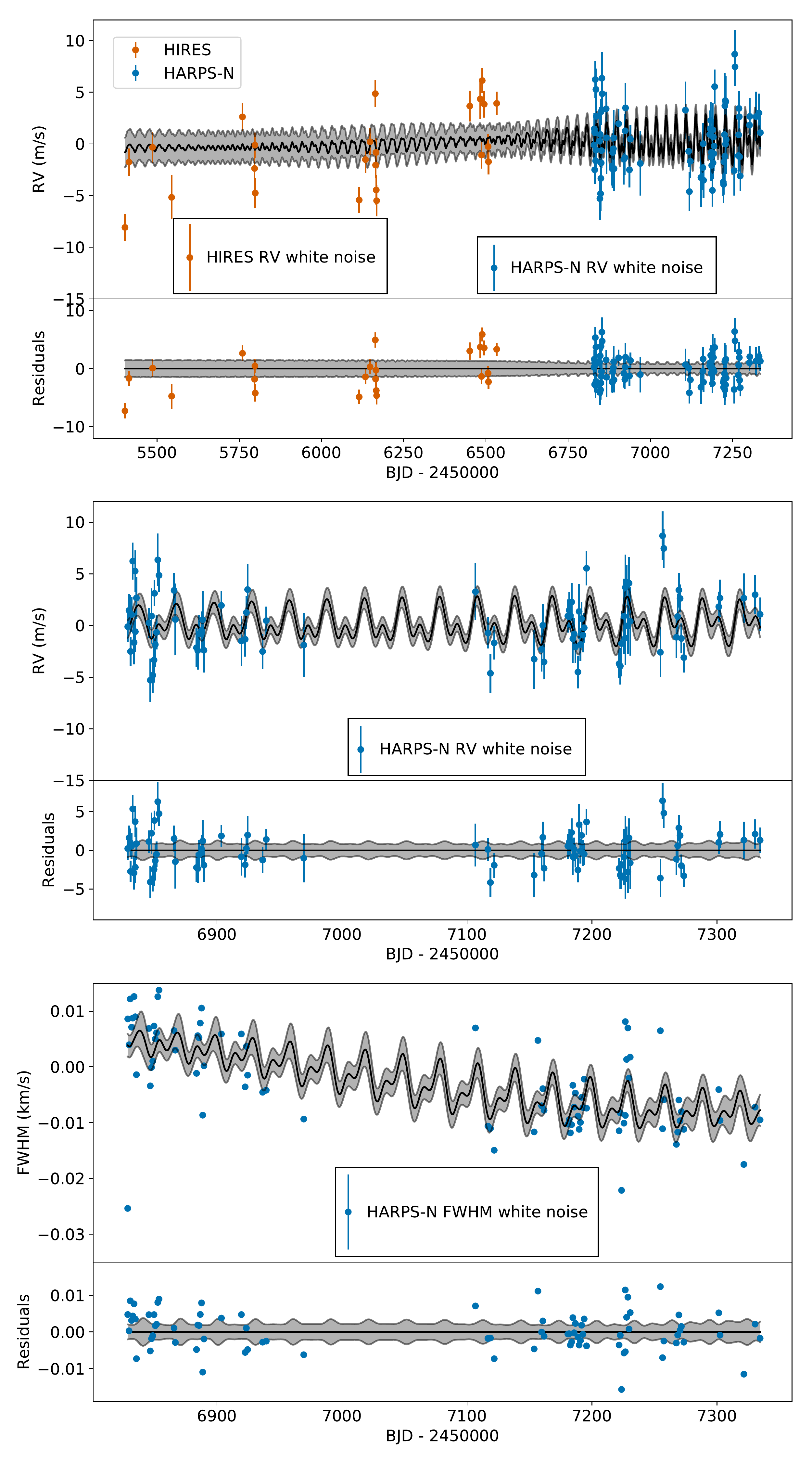}
\caption{Stellar activity and corresponding Gaussian process regression of Kepler-538 (with planetary signal removed). The top subplot shows the HIRES (orange) and HARPS-N (blue) mean-subtracted RV observations and corresponding model fit in the top panel, with residuals in the bottom panel. The black line is the model fit and the gray region is the $1\sigma$ confidence interval (drawn from the full posterior distribution). The data points in boxes correspond to the white noise amplitude modeled for each data set. The middle subplot is a zoom in of the top subplot to the latter two campaigns of observations (only the HARPS-N data). The bottom subplot shows the mean-subtracted FWHM times from HARPS-N (matching the time series of the middle panel) and the corresponding model fit in the top panel, residuals in the bottom panel. Note: two RV data points with error bars greater than $5$ m s$^{-1}$ were removed from the plots (but not the underlying model fit).} \label{stellar_activity_plot}
\end{center}
\end{figure*}

The RV and FWHM observations (and the corresponding model fit) can be seen in Fig.~\ref{stellar_activity_plot}. Additionally, all RV, FWHM, BIS, and $\log R'_{HK}$ values are listed in Table~\ref{table:rv_table}. There is a clear long-term trend in the FWHM observations (and to a lesser extent in the BIS and $\log R'_{HK}$ observations). However, we could not determine whether these trends have a stellar or instrumental origin, nor why there are no similar trends in the RV observations. On the one hand, when we checked three standard stars observed by HARPS-N during the same period of time, only one showed a similar FWHM trend. On the other hand, a FWHM trend in HARPS-N observations was also reported by \citet{benattietal2017} due to a defocusing problem, but that issue was corrected in 2014 March, before our first HARPS-N observations began. Still, perhaps a similar but slower and smaller drift affected our observations.

We first analyzed our observations with a periodogram, then with a correlation plot, and then constructed a model for our spectroscopic data.

\subsubsection{Periodogram Analysis}

\begin{figure*}[ht!]
\epsscale{0.75}
  \begin{center}
      \leavevmode
\plotone{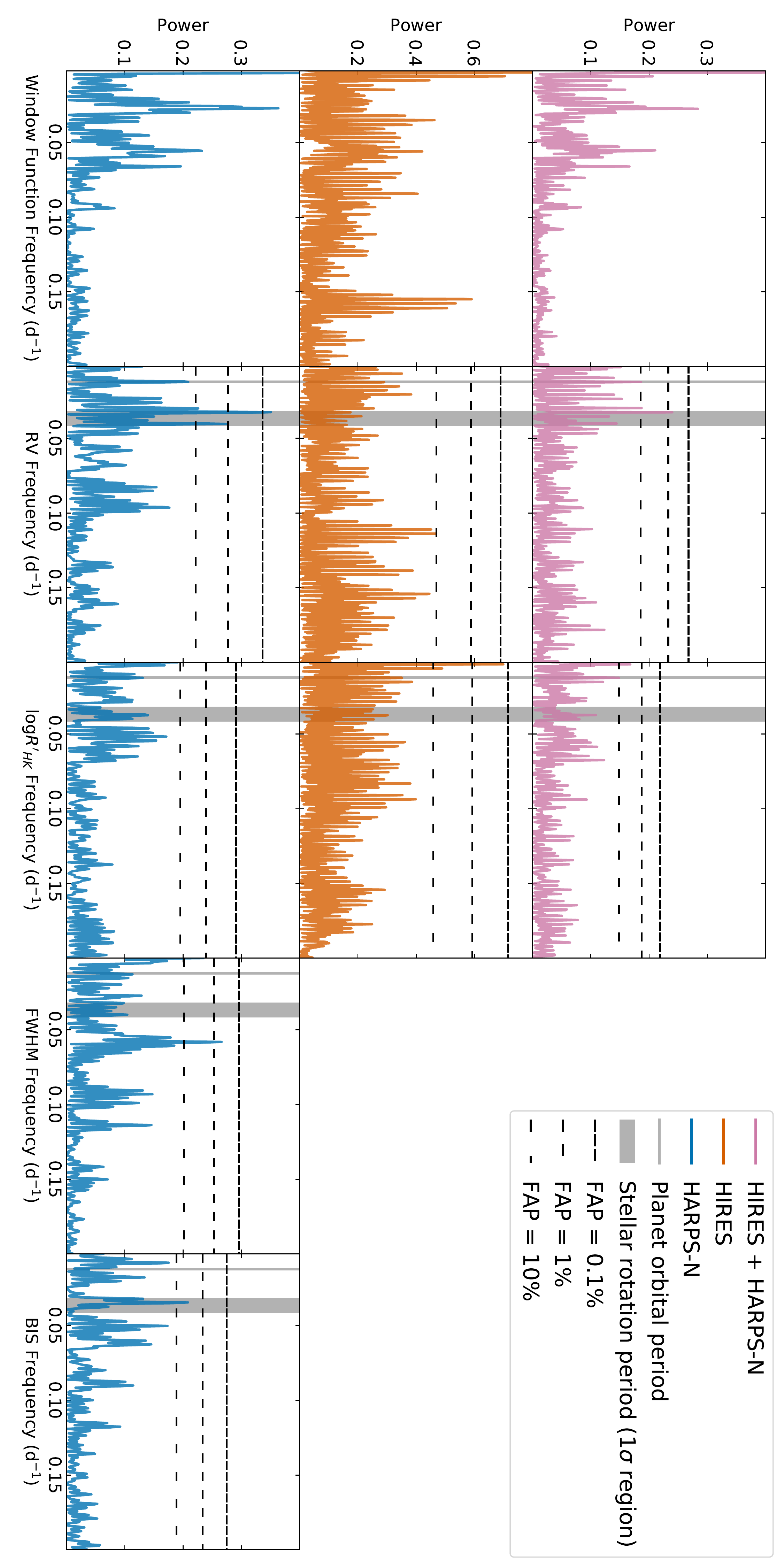}
\caption{Periodograms of the window function (computed from observation times), RV, $\log R'_{HK}$, CCF FWHM, and CCF BIS the Kepler-538 system. Subplots in blue are based on HARPS-N observations, subplots in orange are based on HIRES observations, and subplots in pink are based on both HARPS-N and HIRES. The gray region is the $1\sigma$ confidence interval of the rotation period of Kepler-538 (a stellar activity parameter we estimated in our full model). The gray line is the orbital period of Kepler-538b ($P = 81.74$ days). Lastly, the dashed black lines correspond to various false alarm probabilities.(Note the different y-axis scalings for HIRES.))
} \label{periodogram}
\end{center}
\end{figure*}

Before modeling our spectroscopic observations, we first investigated the frequency structure of our data. We made a generalized Lomb-Scargle periodogram \citep{scargle1982, zechmeisterandkurster2009} of $\log R'_{HK}$, BIS, FWHM, RV, and the window function of the observation time series, all of which can be seen in Fig.~\ref{periodogram}. $\log R'_{HK}$, BIS, and FWHM are indicators of stellar activity (\citealt{quelozetal2001}; also see \citealt[][and references therein]{haywood2015}). The window function shows how the signals are modified by the time sampling of the measurements.

The HARPS-N RV periodogram shows a clear peak at $82$ days, the orbital period of Kepler-538b; the HIRES RV periodogram shows a weaker signal at the same period. None of the other periodograms show a similar feature, lending credence to the RV detection of Kepler-538b. The RV periodograms also exhibit two larger peaks near $0.03-0.04$ days$^{-1}$, interpreted as the rotational frequency. Indeed, as our model fit discussed later in Section~\ref{parameter_estimation} and the results in Table~\ref{results_table} will show, both peaks fall within the $1\sigma$ confidence region of the stellar rotation period. (See Section~\ref{modeling_approach} for a description of our rotation period estimation.) We also find that the long-term trends observed in the activity indices, combined with the spectral window, affect the periodograms, since a long-term trend is clearly noticeable (see Fig.~\ref{stellar_activity_plot} and Table~\ref{table:rv_table}). We removed these trends and found the resulting periodograms show a peak at the rotational period, but nothing at the orbital period.

\subsubsection{Correlation Analysis}

\begin{figure}[ht!]
\epsscale{1.1}
  \begin{center}
      \leavevmode
\plotone{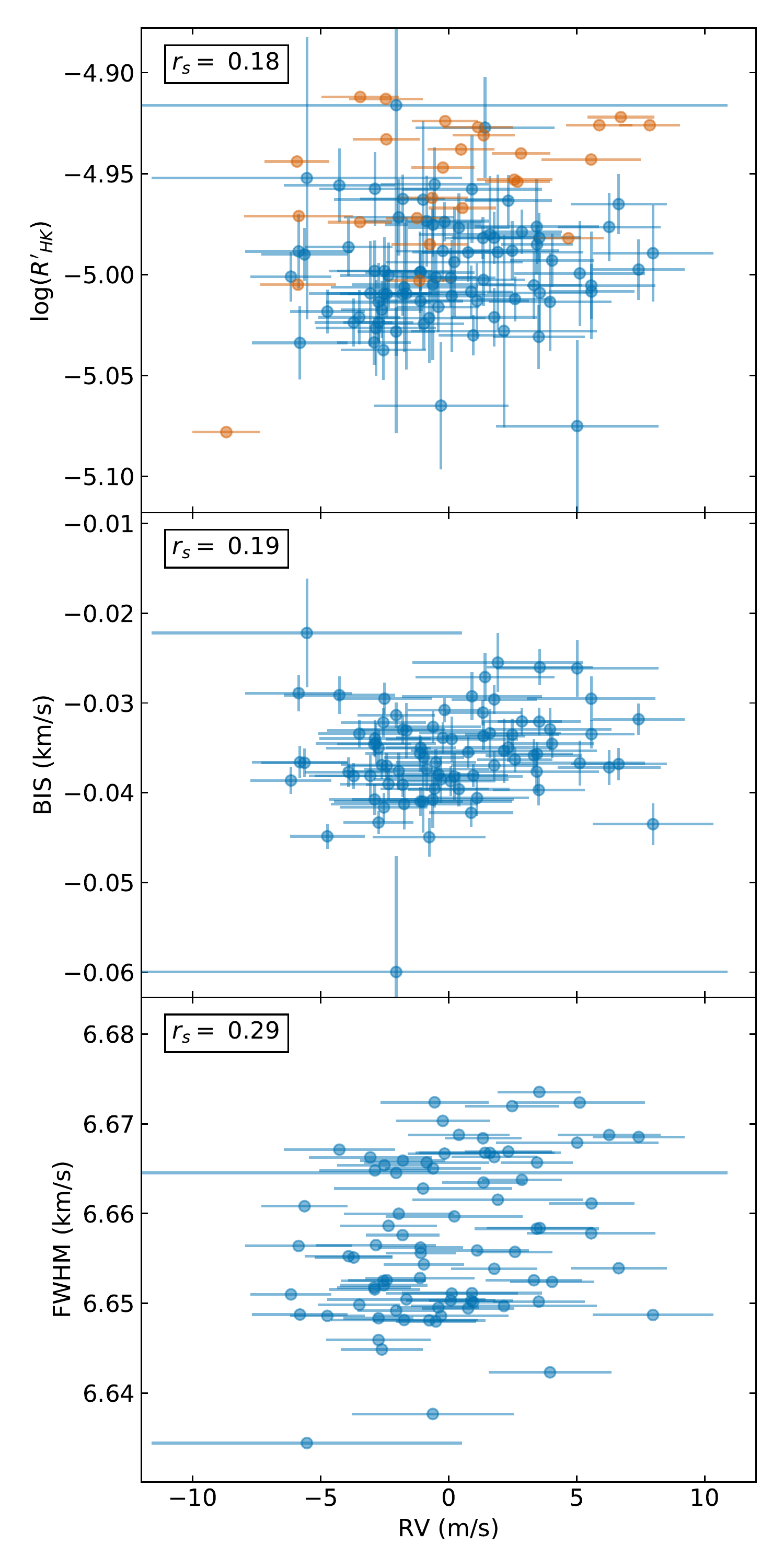}
\caption{Scatter plots of RV vs. $\log R'_{HK}$, BIS, and FWHM for Kepler-538. The RVs have been mean-subtracted and plotted against the other three data types. Blue data points correspond to HARPS-N observations, orange data points to HIRES. In the top left corner of each panel is the Spearman correlation coefficient between the two data sets, an indicator of nonlinear, monotonic correlation. (The coefficients were calculated using the observation values but not their uncertainties.) \label{correlations}}
\end{center}
\end{figure}

We also examined correlations between the RV observations and the other stellar activity indices. As can be seen in Fig.~\ref{correlations}, there is a slightly stronger correlation between RV and FWHM than between RV and BIS or $\log R'_{HK}$. However, there may also be useful information in the correlations between RV and BIS or $\log R'_{HK}$. In order to test this, we cross-checked results that included BIS and $\log R'_{HK}$ in the modeling against those that did not and found consistent results. For this reason, and for the sake of simplicity, in this paper we only report our analysis of RVs in conjunction with FWHM observations.

\subsubsection{General RV Modeling Approach}\label{modeling_approach}
In order to model our RV and FWHM observations, we followed the method described in \citet[][hereafter R15]{rajpauletal2015}, which establishes a method to characterize stellar activity that uses simultaneous regression of distinct data types (with potentially distinct time series). Here we briefly discuss Gaussian process (GP) regression and the novel approach to GPs used by R15.

In brief, a GP is a stochastic process that captures the covariance between observations and allows for the modeling of correlated noise \citep{rasmussenandwilliams2006}. A GP is specified by a covariance matrix in which the diagonal elements are the individual observation variances and each off-diagonal element describes the covariance between two observations. The values of the off-diagonal elements are determined by a kernel function, which describes the nature of the correlated noise. GPs provide a great deal of flexibility that has made them an effective tool to account for stellar activity \citep{haywoodetal2014}. R15 recommended characterizing stellar activity with a quasi-periodic (QP) kernel, which balances physical motivation with simplicity. The QP kernel uses four parameters (commonly called hyperparameters) and defines the covariance matrix as follows:

\begin{multline} \label{eqn:QP}
    K_{\mathrm{QP}}(t_i,t_j) = \\ h^2\exp{\bigg(-\dfrac{\sin^2{\big(\pi(t_i-t_j)/P_*\big)}}{2\lambda^2_p}-\dfrac{(t_i-t_j)^2}{2\lambda_e^2}\bigg),}
\end{multline}

where $t_i$ and $t_j$ are observations made at any two times, $h$ is the amplitude hyperparameter (though not a \textit{true} amplitude, as it incorporates some multiplicative constants), $P_*$ is the period of the variability (i.e., the rotation period in the case of stellar activity), $\lambda_p$ is the inverse harmonic complexity (a smoothness factor that acts as a proxy for the number of turning points and inflection points per rotation period), and $\lambda_e$ is an exponential decay factor (scaling with, though not exactly equal to, the decay timescale of the spots on the star).

One of the key insights of R15 is the way in which they related multiple GPs to one another. GP regression can be used on multiple data sets by constructing a covariance matrix that describes the covariances between two observations of any type. In our case, this means any possible pairing of RV-RV, RV-FWHM, or FWHM-FWHM data points. The following equations (based on equations 13 and 14 from R15) relate RV and FWHM:

\begin{equation} \label{eqn:13}
    \Delta RV = V_cG(t) + V_r\dot{G}(t)
\end{equation}
\begin{equation} \label{eqn:14}
    FWHM = F_cG(t)
\end{equation}

Here, $G(t)$ is an underlying GP directly quantifying stellar activity, and $V_c$, $V_r$, and $F_c$ are amplitude parameters corresponding to the RV convective blueshift suppression effect, RV rotation modulation, and FWHM signal amplitude (note that this means there are three amplitude parameters instead of the single $h$ parameter expressed in Equation~\ref{eqn:QP}). Because RVs and FWHMs respond differently to the underlying stellar activity, this approach allows for more rigorous characterization of the stellar activity than methods using only RV observations, which improves the separation of the stellar and planetary signals.

We followed R15 and simultaneously modeled the HIRES RV data as well as the HARPS-N RV and FWHM data. This included a separate offset parameter and noise parameter (added in quadrature to the uncertainties) for both RV data sets and the FWHM data set (for a total of three offset parameters and three white noise parameters). Finally, the RV reflex motion due to the planet was characterized by a simple five-parameter orbital model: reference epoch, orbital period, reflex motion semi-amplitude, eccentricity, and longitude of periastron.

Because we conducted a joint fit to both photometry and spectroscopy, all orbital parameters except for reflex motion semi-amplitude are simultaneously used in our photometric model. In other words, reference epoch, orbital period, eccentricity, and longitude of periastron are used in both the photometric and spectroscopic components of our full model.

For all of the parameters used in the spectroscopic portion of the model, we assumed uniform, Jeffreys, or modified Jeffreys priors. The specific types and bounds of the priors are all listed in Table~\ref{results_table}.

\subsection{Parameter Estimation} \label{parameter_estimation}

Overall, our full model included a photometric baseline offset parameter, a photometric white noise parameter, two quadratic limb-darkening parameters, the impact parameter, the transit duration, the planet radius relative to the stellar radius, the reference epoch, the orbital period, eccentricity, longitude of periastron, the reflex motion semi-amplitude, three spectroscopic offset parameters and three spectroscopic white noise parameters (for HIRES RV, HARPS-N RV, and HARPS-N FWHM), and six GP hyperparameters (two corresponding to the two RV semi-amplitudes and one corresponding to the FWHM semi-amplitude in Equations~\ref{eqn:13} and~\ref{eqn:14}, as well as the stellar rotation period, a smoothness factor, and an exponential decay factor). This yielded a total of 24 parameters, all of which are also listed in Table~\ref{results_table}. (Note: because we only modeled the detrended and flattened photometry, our estimation of the stellar rotation period was derived solely from our spectroscopic data.)

We estimated model parameters using \texttt{MultiNest} \citep{ferozetal2009,ferozetal2013}, a Bayesian inference tool for parameter space exploration, especially well suited for multimodal distributions. We used the following \texttt{MultiNest} settings for our parameter estimation: constant efficiency mode, importance nested sampling mode, multimodal mode, sampling efficiency = 0.01, 1000 live points, and evidence tolerance = 0.1.

\begin{figure}
\epsscale{1.2}
  \begin{center}
      \leavevmode
\plotone{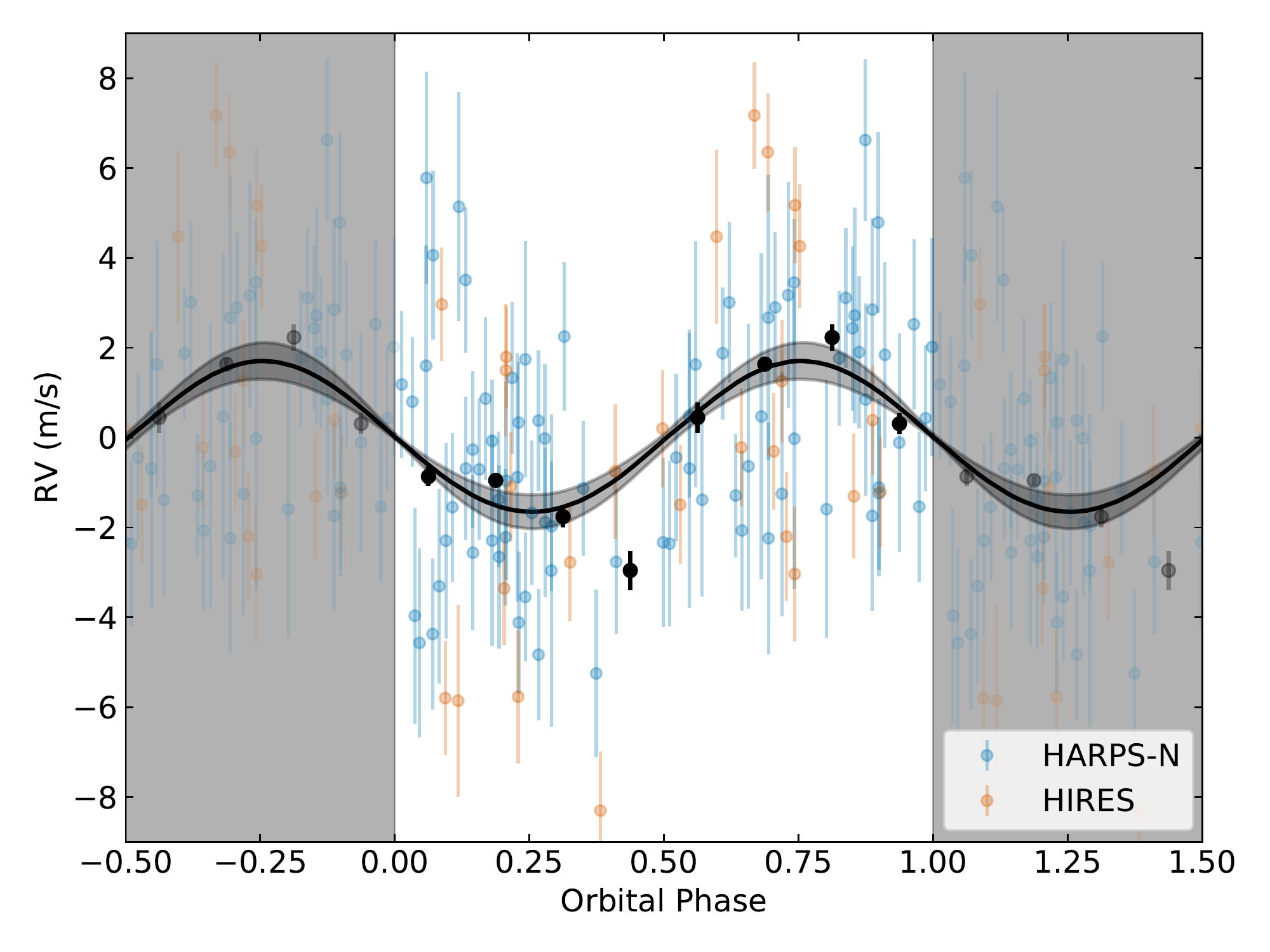}
\caption{Kepler-538 RVs (with stellar activity subtracted) as a function of the orbital phase of Kepler-538b. Observations from HARPS-N and HIRES are plotted in blue and orange respectively, and binned data points are plotted in black. Data in the gray regions on each side of the plot are duplicates of the data in the white region. The median model and $1\sigma$ confidence interval are plotted as a black line and gray region respectively. Note: two RV data points with error bars greater than $5$ m s$^{-1}$ were removed from the plot (but not the underlying model fit).} \label{rv_plot}
\end{center}
\end{figure}

Our full results from this analysis are presented in Table~\ref{results_table} and discussed in Section~\ref{results_discussion}. Further, the best-fit transit model is plotted against the photometric data in Fig.~\ref{transit_plot} and the phase-folded, stellar-activity-removed RV observations and model are presented in Fig.~\ref{rv_plot}. We find Kepler-538b to have a mass of $M_p = 10.6^{+2.5}_{-2.4} M_\oplus$, a radius of $R_p = 2.215^{+0.040}_{-0.034} R_\oplus$, a mean density of $\rho_p = 0.98 \pm 0.23$ $\rho_\oplus$, and negligible eccentricity (consistent with zero, $< 0.11$ at $95\%$ confidence). Notably, thanks to the \textit{Gaia} parallax, our uncertainty on the planetary radius is less than 2\%. For context, the average uncertainty, $0.037 R_\oplus$, is only $236$ km, approximately the distance between Portland and Seattle\footnote{\url{https://www.distancecalculator.net/from-portland-to-seattle}}. Finally, we also note that our estimates of transit parameters are all within $1\sigma$ of those reported in the original Kepler-538b validation paper \citep{mortonetal2016}.

\subsection{Model Tests}

In order to confirm the validity of the results from our RV analysis, we conducted a number of tests designed to verify both our method of analysis and its output. These tests included removing our prior knowledge (obtained via transit photometry) of the transit time and period, injecting and recovering synthetic planet signals into the RV data, and removing the GP to model only the planet signal.

\subsubsection{Removing the Transit Prior}

The first test we conducted was to repeat our analysis without any photometric observations, thereby removing the strong photometric constraints on the transit time and orbital period. We refit our model with a prior of BJD-2453833 = Unif($172$,$252$) on transit time, $P$ = Jeffreys(40,120) on orbital period, and the same priors on all other parameters that we previously used in our full analysis. We fit against only RV and FWHM observations, so we did not have any photometric parameters. Our choice of transit time prior was large enough to be naive, but small enough to exclude other transit times modulo some number of orbital periods. Similarly, our choice of orbital period prior was large enough to be naive, but small enough (on the lower end) to prevent overlap with the stellar rotation period of $25-30$ days.

The results were consistent with the full simultaneous fit to spectroscopy and photometry. Of course, the posterior distributions on transit time and orbital period were much wider, which is to be expected. Specifically, the transit time was found to be $t_0$  (BJD - 2454833) $= 203^{+14}_{-13}$ and the period was found to be $P = 82.25^{+0.62}_{-0.74}$ days. However, all parameters agreed within $1\sigma$ of those from the full, simultaneous fit results. Further, all uncertainties (other than those of transit time, period, and eccentricity) were of a similar scale to those from the full model.

\subsubsection{Injection Tests}

The next test we conducted was to introduce a $1.7$ m s$^{-1}$, non-eccentric, sinusoidal planetary signal into the RV data at various periods to see whether the signal could be recovered, whether the measured RV semi-amplitude was accurate, and whether the uncertainties were similar to those for Kepler-538b. We ran four separate model fits with a synthetic planetary signal introduced at $60$ days, $70$ days, $90$ days, and $100$ days respectively. For each data set, we modeled Kepler-538b and the synthetic signal simultaneously, including eccentricity in the model for both planets. To reduce computational expenses, we did not model the \Kepler\ photometry for these tests, instead we applied Gaussian priors to the orbital period and transit time of Kepler-538b based on the values from our main results (see Table~\ref{results_table}). As for our injected signal, we applied Gaussian priors to transit time and orbital period, centered respectively on the transit time and orbital period of the injected signal, with the same variance on transit time and same fractional variance on orbital period as for Kepler-538b. Finally, priors on semi-amplitude, eccentricity, and longitude of periastron were identical to those for Kepler-538b.

In all four model fits, we recovered the semi-amplitude of the injected signal to within $1\sigma$ of $1.7$ m s$^{-1}$ (except for the 60d injection test, for which we found a semi-amplitude that was less than $1.7$ m s$^{-1}$ by $1.1\sigma$). Further, the recovered semi-amplitude uncertainty of the injected planets were all on the order of $0.4-0.5$ m s$^{-1}$, similar to the error bars on the semi-amplitude of Kepler-538b. Finally, in all four cases, the measured eccentricity of the injected planet was consistent with zero to within $2\sigma$.

\subsubsection{Fitting without a GP}

Another important test we conducted was trying to model the RVs of Kepler-538b without accounting for the stellar activity at all. We did this by simply running the analysis without the GP. If the GP regression adequately accounted for the stellar activity (rather than subsume and weaken the planetary signal), we would expect to recover a similar RV semi-amplitude for the planet when the GP is excluded, as well as either comparable or larger uncertainties.

And this is indeed what we find. Without a GP, we found an RV semi-amplitude of $K = 2.06^{+0.49}_{-0.46}$ m s$^{-1}$, within $1\sigma$ of the semi-amplitude found when a GP was included. Similarly, all other parameters in common between the two model fits agreed to within $1\sigma$, adding confidence to our results.

This particular test illustrates that our choice to use a GP to account for stellar activity was sufficient for this system and data set, though not strictly necessary. This may be due to the long evolution time scale of the stellar activity and the large difference in periods between stellar rotation and planetary orbital period. However, we cannot rely on favorable stellar features in general, therefore it is best to err on the side of caution and use a sufficiently sophisticated method (e.g. GP regression) to characterize stellar activity signals.

\begin{figure}
\epsscale{1.3}
  \begin{center}
      \leavevmode
\plotone{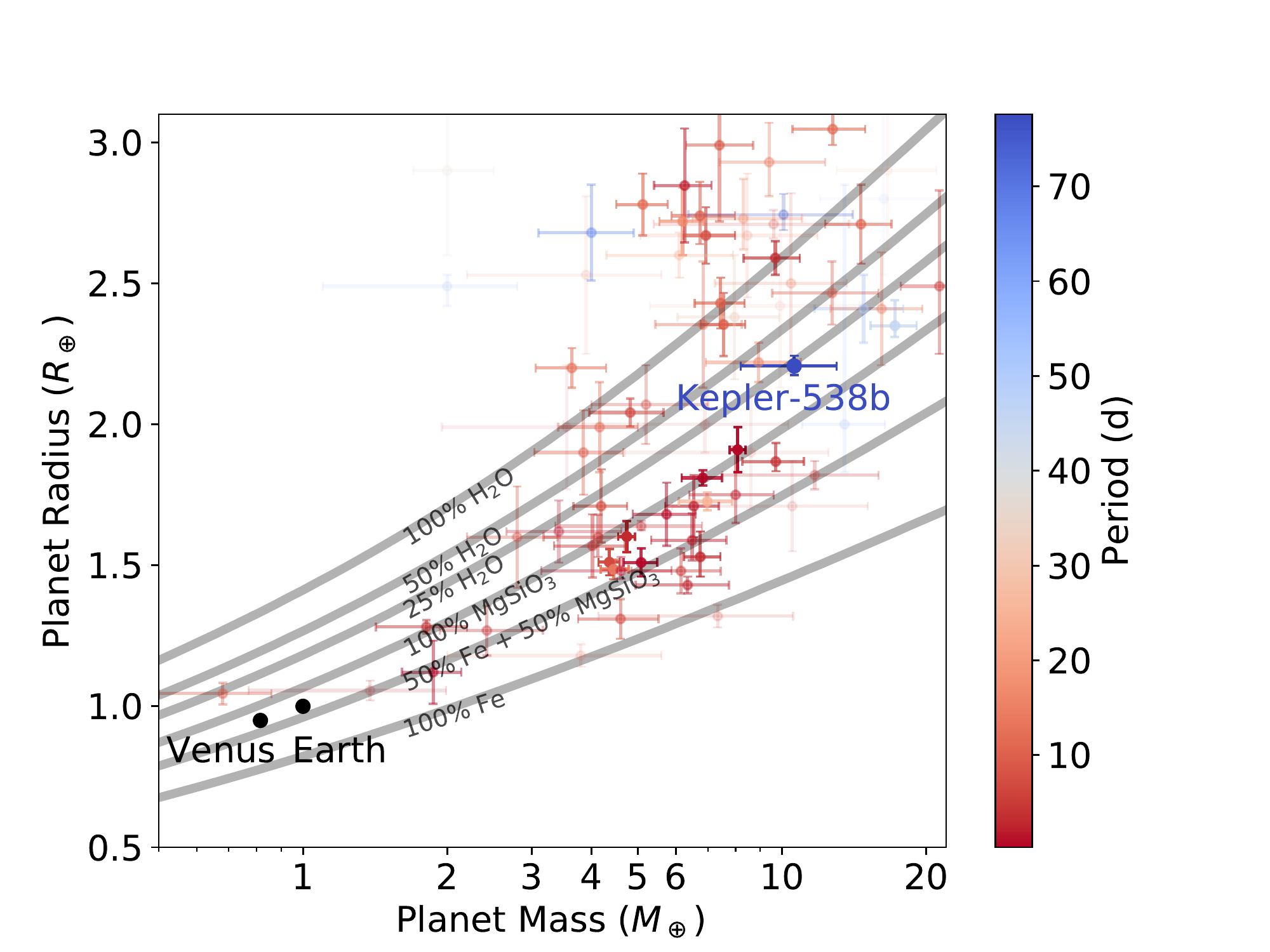}
\caption{Mass-radius diagram of transiting planets with fractional mass and radius uncertainties less than 50\%. Planet colors correspond to orbital period, with short periods in red and long periods (such as Kepler-538b) in blue. Further, except for Kepler-538b, planets with larger fractional mass and radius uncertainties are fainter. Venus and Earth are also labeled and plotted in black for reference. Gray lines correspond to planetary compositions (from top to bottom) of 100\% H$_2$O, 50\% H$_2$O, 25\% H$_2$O, 100\% MgSiO$_3$, 50\% MgSiO$_3$ + 50\% Fe, and 100\% Fe, respectively \citep{zengandsasselov2013,zengandsasselov2016}. Kepler-538b lies closest to the 25\% H$_2$O composition line. The planet likely consists of a significant fraction of ices (dominated by water ice), in addition to rocks/metals, and a small amount of gas.} \label{mr_plot}
\end{center}
\end{figure}

\section{Results and Discussion} \label{results_discussion}

The results of our stellar characterization and light curve, RV, and FWHM modeling can be found in Tables~\ref{stellar_table} and~\ref{results_table}.

After conducting our model fits and running the requisite follow-up tests, we found the mass of Kepler-538b to be $M_p = 10.6^{+2.5}_{-2.4} M_\oplus$. Combining this with the planetary radius of $R_p = 2.215^{+0.040}_{-0.034} R_\oplus$ resulted in a planetary density of $\rho_p = 0.98 \pm 0.23$ $\rho_\oplus$, or $5.4 \pm 1.3$ g cm$^{-3}$. 

Owing to its long orbital period, and its location on the mass-radius diagram, Kepler-538b likely consists of a significant fraction of ices (dominated by water ice), in addition to rocks/metals, and a small amount of gas \citep{zengetal2018}. Its host star is slightly less massive than our own Sun. Because the luminosity of a main-sequence star is a strong function of its mass (typically to the power of 3 or 4), the luminosity of the host star Kepler-538 is somewhat less than the Sun. Therefore, the snowline in the disk when this system was formed was closer in, increasing the likelihood for Kepler-538b to accrete ices during its formation. 

The estimated bulk density of Kepler-538b is comparable to that of the Earth. However, this high mean density is partly due to its high mass resulting in more compression of materials under self-gravity. Its uncompressed density, as revealed by the mass-radius curves \citep{zengandsasselov2013,zengandsasselov2016} in Fig.~\ref{mr_plot}, is consistent with a composition somewhat less dense than pure-rocky and/or Earth-like rocky (1:2 iron/rock mixture). One ready explanation is that Kepler-538b is an icy core, which for some reason had not accreted as much gas as our own Uranus or Neptune (both are estimated to have a few up to ten perfect mass of gas). 

The eccentricity of Kepler-538b is small (less than $0.11$ with 95\% confidence). However, the planet may still have arisen from a dynamical origin, that is, inward planet migration due to planet-planet gravitational interactions \citep{raymondetal2009}. Some planet formation theories have suggested the formation of multiple icy cores in relatively adjacent space near the snowline around a host star, increasing the likelihood of dynamical interactions among them and resulting in inward scatterings for some of them. If Kepler-538b were scattered inward, then its orbital eccentricity could have been higher initially, and then damped to its current value through interactions with the disk when the disk was still around. Alternatively, inward migration through planet-disk interactions may be a more likely scenario, since a disk would always keep the planet orbital eccentricity low \citep{chambers2018, morbidelli2018} and would probably be required to damp any eccentricity from scattering.

In summary, Kepler-538b is only the “tip of a huge iceberg”, likely representing a class of planets common in our Galaxy, but which are not found in our own solar system. The absence of planets in between the size of the Earth and Neptune (about four Earth radii) is linked to the formation/presence of a gas giant -- Jupiter \citep{izidoroetal2015,barbatoetal2018}, and vice versa.

To date, very few exoplanets have been found on long-period orbits that also have any kind of mass measurements. In fact, according to the NASA Exoplanet Archive\footnote{\url{https://exoplanetarchive.ipac.caltech.edu/}. This number was determined by constraining orbital period $>50$ days, planet mass $<11 M_{\mathrm{Jup}}$, planet mass limit flag = 0 (to remove upper limit results), planet circumbinary flag = 0, planet transit flag = 1, and planet RV flag = 1.} (accessed 2019 July 31), there are only 10 transiting exoplanets (excluding Kepler-538b) with an RV mass measurement and an orbital period greater than 50 days. If we look at other common methods of mass measurement (specifically transit timing variations and dynamical mass measurements of circumbinary planets), that number only increases to 37.

\begin{figure}[ht!]
\epsscale{1.2}
  \begin{center}
      \leavevmode
\plotone{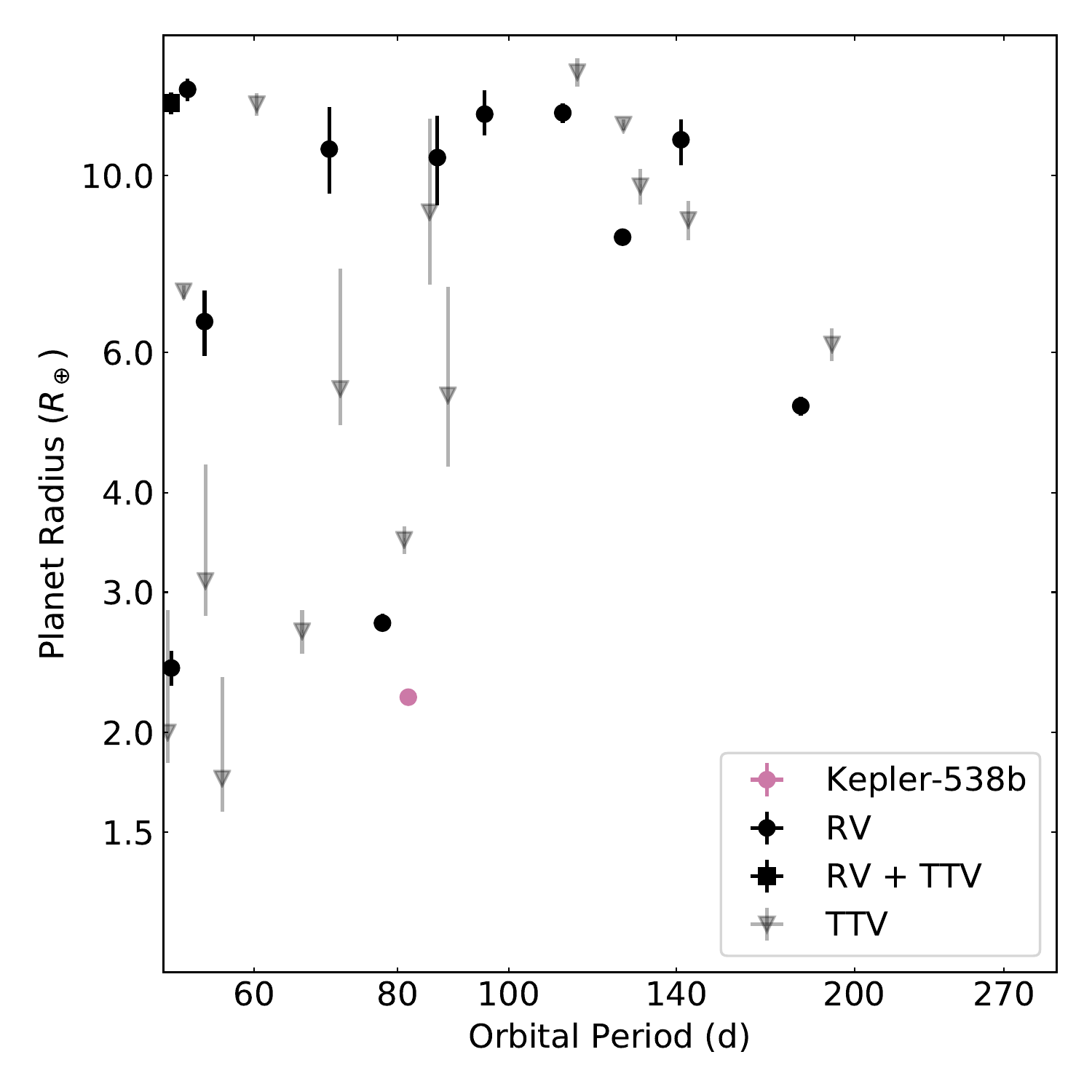}
\caption{Orbital period versus planet radius for all transiting exoplanets with $P > 50$ days and RV or transit timing variation (TTV) mass measurements. Data for all planets besides Kepler-538b were retrieved from the NASA Exoplanet Archive (accessed 2019 February 16). Kepler-538b is plotted as a pink circle, all other exoplanets with RV mass measurements are plotted as black circles, one exoplanet (Kepler-117c) has a jointly derived mass from RV and TTV measurements and is plotted as a black square, and exoplanets with only TTV mass measurements are plotted as gray triangles. (Period and radius uncertainties are plotted for all planets, including Kepler-538b, but are smaller than the data points in many cases.) At long periods ($P>50$ days), Kepler-538b is the smallest transiting exoplanet with an RV mass measurement, and Kepler-20d is the only such planet with a lower mass (by $0.5 M_\oplus$). Overall, there are very few mass measurements for planets in the long-period, small-radius regime of Kepler-538b.} \label{per_v_rad}
\end{center}
\end{figure}

Further, most of those planets are quite large, more similar to Jupiter or Saturn in mass and radius than Neptune or Earth. Fig.~\ref{per_v_rad} demonstrates where Kepler-538b fits into this sparse region of parameter space. Kepler-538b is one of the very few small, low-mass planets well characterized to date. 

As the sample of small, long-period planets with precisely determined masses and densities grows, we will be able to address a number of fundamental questions. For example, what effect does stellar incident flux have on the size and composition of exoplanets? Since most known exoplanets have periods shorter than that of Mercury, it is difficult to analyze exoplanet composition and size for incident fluxes comparable to or less than that of Earth. Similarly, is there a relationship between the location or depth of the planet radius occurrence gap detected by \citet{fultonetal2017} and a planet's mass or composition? Further characterization of this gap at longer periods would help confirm (or refute) the photoevaporation explanation of the gap and therefore provide insights about exoplanet formation.

\subsection{Detection of Kepler-538b with Other Methods}

As methods of detecting exoplanets become more sensitive, regions of parameter space accessible to multiple detection methods will grow, and with them the opportunity to more rigorously characterize the planet population and calibrate detection methods against one another. Kepler-538b pushes RV characterization further into the low-mass, long-period planet regime. As a result, it is interesting to explore whether other methods might also be able to characterize such a planet.

To begin with, there is no possibility of detecting an astrometric signal of Kepler-538b. \citet{perrymanetal2014}, which analyzed the expected planet yield from \textit{Gaia} astrometry, found that the expected along-scan accuracy per field of view for \textit{Gaia} would be $\sigma_{\mathrm{fov}} = 34.2 \mu$as for a star like Kepler-538 ($G = 11.67$). While they required an astrometric signal of $3\sigma_{\mathrm{fov}}$ for a detection, the astrometric signal of Kepler-538b is only $0.095 \pm 0.022 \mu$as, over 1000 times smaller than this detection threshold.

Similarly, a planet like Kepler-538b is very unsuitable for direct imaging. According to the NASA Exoplanet Archive\footnote{\url{https://exoplanetarchive.ipac.caltech.edu/}} (accessed 2019 July 28), there are no directly imaged planets less massive than $2 M_{\mathrm{Jup}}$ or closer to their host star than $2$ au, both of which disqualify Kepler-538b. Further, direct imaging is well suited for young stars which still host self-luminous planets, but the median estimated age of Kepler-538 is $3.8$ Gyr, older than nearly every host star of a directly imaged planet on the NASA Exoplanet Archive (there are only two exceptions, WISEP J121756.91+162640.2 A and Oph 11).

Unlike astrometry and direct imaging, \citet{pennyetal2019} determined that a planet with the mass and semi-major axis of Kepler-538b would be just inside the microlensing sensitivity curve of the \textit{Wide Field Infrared Survey Telescope (WFIRST)}. They estimated that if every star hosted a planet like Kepler-538b, we could expect \textit{WFIRST} to detect a microlensing signal from roughly 10-30 such planets during the course of the full mission (see Fig. 9 from \citealt{pennyetal2019}).

\subsection{Potential for Atmospheric Characterization}

\begin{figure}[ht!]
\epsscale{1.15}
  \begin{center}
      \leavevmode
\plotone{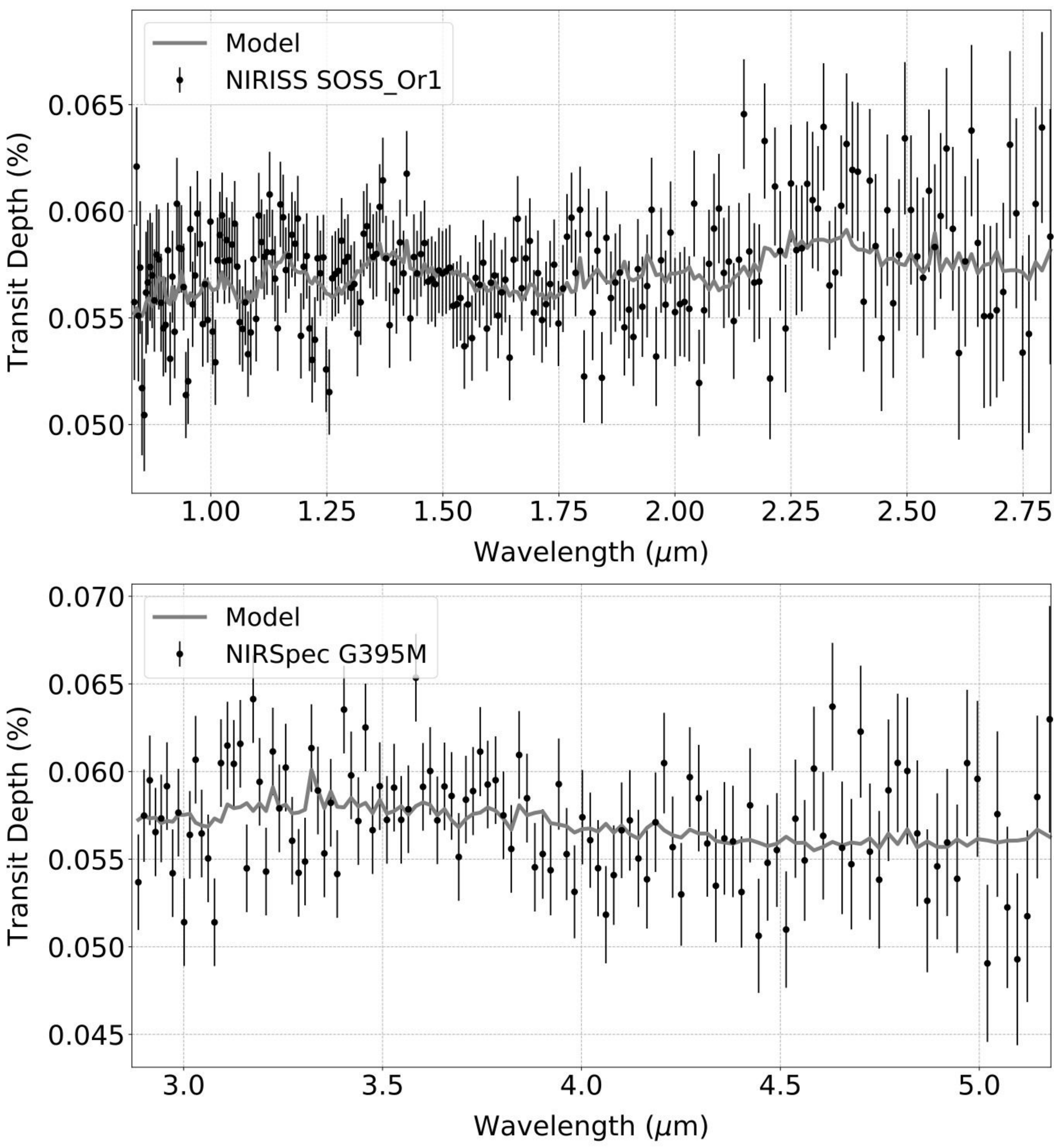}
\caption{A simulated transmission spectrum of Kepler-538b with five transits observed with \textit{JWST}. The model spectrum, with low metallicity (five times solar) and no clouds, is shown as a gray line. The black data points are the simulated instrument spectra, using NIRISS SOSS-Or1 (0.81-2.81 $\mu$m) and NIRSpec G395M (2.87-5.18 $\mu$m).} \label{pandexo}
\end{center}
\end{figure}

One interesting question to ask about Kepler-538b is whether or not it may be amenable to atmospheric characterization via transmission spectroscopy. The \textit{James Web Space Telescope} (\textit{JWST}; \citealt{gardneretal2006,demingetal2009,kalirai2018}) will devote a significant portion of its mission to the characterization of exoplanet atmospheres. The spectra shown in Fig.~\ref{pandexo} for the atmosphere of Kepler-538b were generated by the \textit{JWST} Exoplanet Targeting (\texttt{JET}) code (C. D. Fortenbach \& C. D. Dressing 2019, in preparation) assuming five observed transits. This code first takes the observed planet and system parameters ($R_p$, period, insolation flux, $R_*$, $T_{eff}$, and $J$-band magnitude) and then derives other key parameters (semi-major axis, $T_{\mathrm{eq}}$, planet surface gravity, planet mass, and transit duration). In this case we used the planet mass already determined in this paper. We also assumed an optimistic low-metallicity (five times solar) planetary atmosphere with no clouds. \texttt{JET} then used \texttt{Exo-Transmit} \citep{kemptonetal2017} to generate model transmission spectra and used \texttt{Pandexo} \citep{batalhaetal2017} to generate simulated instrument spectra. We focused on the Near InfraRed Imager and Slitless Spectrograph (NIRISS) SOSS-Or1 and NIRSpec G395M instruments/modes since they are, according to \citet{batalhaandline2017}, best suited for exoplanet transmission spectroscopy. Finally, the \texttt{JET} code performed a statistical analysis for multiple transits and determined if the simulated instrument spectra fit the model well enough to confirm a detection. Given current estimates of the precision (noise floor) of these \textit{JWST} instruments (as well as visual inspection of the simulated spectra after five transits in Fig.~\ref{pandexo}), it would likely be very difficult to detect the Kepler-538b atmosphere even with a large number of transit observations with \textit{JWST}.

Perhaps other next-generation observatories such as the Thirty Meter Telescope \citep{sanders2013}, the Extremely Large Telescope \citep{udryetal2014}, the Giant \textit{Magellan} Telescope \citep{johnsetal2012}, or the \textit{Large UV/Optical/IR Surveyor} \citep{luvoirteam2018} will be able to make such a project feasible.

\subsection{Possibility of a Second Planet in the System} \label{cand2}

Some early versions of the \Kepler\ catalog included a weak transit signal at $117.76$ days and labeled it as a planet candidate (K00365.02). However, one early catalog instead labeled it as a false positive \citep{mullallyetal2015} and the final \Kepler\ catalog \citep[DR25;][]{thompsonetal2018} did not detect a candidate at that period at all (or even a threshold crossing event, the broadest detection category in the \Kepler\ pipeline). Further, the \Kepler\ False Positive Working Group \citep{brysonetal2017} investigated K00365.02 and could not determine a final disposition; they did however flag the candidate with a ``Transit Not Unique False Alarm'' flag, meaning ``the detected transit signal is not obviously different from other signals in the flux light curve.''\footnote{\url{https://exoplanetarchive.ipac.caltech.edu/\\docs/API\_fpwg\_columns.html}}

The radius of K00365.02 was reported on the NASA Exoplanet Archive as $0.62^{+0.10}_{-0.03} R_\oplus$. Assuming a pure iron composition and using \citet{zengandsasselov2013} and \citet{zengandsasselov2016} yields an upper limit mass of $0.37^{+0.25}_{-0.05} M_\oplus$ and an upper limit semi-amplitude of $5.3^{+3.4}_{-0.8}$ cm s$^{-1}$, well below the detection threshold for HARPS-N, HIRES, or any other spectrograph. However, for the sake of rigor, we also ran a two planet model for Kepler-538b and K0035.02 on our RV and FWHM data (similar to our main model). Instead of jointly modeling photometry, we applied period and transit time priors on Kepler-538b and K00365.02 (the former based on our final results, the latter determined from the NASA Exoplanet Archive\footnote{\url{https://exoplanetarchive.ipac.caltech.edu/}}; accessed 31 July 2019). Our results showed an RV semi-amplitude at $117.76$ days of $K = 0.26^{+0.28}_{-0.18}$ m s$^{-1}$, negligible and consistent with zero at less than $1.5\sigma$. 

Additionally, the periods of Kepler-538b and K00365.02 are not in or near a first-order mean motion resonance (or second-order, for that matter), so we do not expect a large, detectable transit timing variation (TTV) signal on Kepler-538b either \citep{lithwicketal2012}. Indeed, the NASA Exoplanet Archive (accessed 31 July 2019) does not report a TTV flag for Kepler-538b. As a result, with an unverified transit signal, a negligible RV signal, and an apparently negligible TTV signal, the existence of K00365.02 remains inconclusive.

\section{Summary and Conclusions} \label{conclusion}

In this paper, we analyze the Kepler-538 system in order to determine the properties of Kepler-538b, the single, known exoplanet in the system. Kepler-538 is a $0.924 M_\odot$, G-type star with a visual magnitude of $V = 11.27$. We model the \Kepler\ light curve and determine the orbital period of Kepler-538b to be $P = 81.74$ days and the planetary radius to be $R_p = 2.215^{+0.040}_{-0.034} R_\oplus$ (for reference, $0.037 = 236$ km, approximately the distance between Portland and Seattle\footnote{\url{https://www.distancecalculator.net/from-portland-to-seattle}}). These results are in agreement with previous transit fits. We also determine the planetary mass by accounting for stellar activity via a GP regression that uses information from the FWHM and RV observations simultaneously. Our model fit yields a mass estimate for Kepler-538b of $M_p = 10.6^{+2.5}_{-2.4} M_\oplus$. Combined, these results show the planet to have a density of $\rho_p = 0.98 \pm 0.23 \rho_\oplus = 5.4 \pm 1.3$ g cm$^{-3}$. This suggests a composition and atmosphere somewhere between that of Earth and Neptune, with a significant fraction of ices (dominated by water ice), in addition to rocks/metals, and a small amount of gas \citep{zengetal2018}.

To date, there have been very few precise and accurate mass measurements of long-period exoplanets. Beyond 50 days, Kepler-538b is only the 11th transiting exoplanet with an RV mass measurement (NASA Exoplanet Archive\footnote{\url{https://exoplanetarchive.ipac.caltech.edu/} This number was determined by constraining Orbital Period $>50$ days, Planet Mass $<11 M_{\mathrm{Jup}}$, Planet Mass Limit Flag = 0 (to remove upper limit results), Planet Circumbinary Flag = 0, Planet Transit Flag = 1, and Planet RV Flag = 1.}; accessed 2019 May 4). Additional, well-constrained mass measurements of long-period planets will improve our understanding of the long-period exoplanet population. Beyond that, they will also help to answer questions about the short-period planet population, such as the nature of the planetary radius occurrence gap \citep{fultonetal2017} and the effect of stellar flux on exoplanet compositions and atmospheres.

With new, next-generation spectrographs such as HPF \citep{mahadevanetal2010,mahadevanetal2014}, KPF \citep{gibsonetal2016, gibsonetal2018}, EXPRES \citep{jurgensonetal2016}, ESPRESSO \citep{megevandetal2010}, and NEID \citep{schwabetal2016} coming online now or in the near future, our ability to characterize long-period exoplanets will only improve. Better data will require more advanced analysis methods to extract as much information as possible. The methods used in this paper, such as GP regression, injection tests, and simultaneous modeling of RV observations and stellar activity indices, are valuable tools that strengthen the analysis of spectroscopic data, improve exoplanet characterization, and therefore better our understanding of the exoplanet population as a whole.

\acknowledgments
\acknowledgments
A.W.M. is supported by the NSF Graduate Research Fellowship grant No. DGE 1752814.

V.M.R. thanks the Royal Astronomical Society and Emmanuel College, Cambridge, for financial support.

C.D.D. acknowledges support from the NASA K2 Guest Observer program through grant 80NSSC19K0099.

This work was performed in part under contract with the California Institute of Technology (Caltech)/Jet Propulsion Laboratory (JPL) funded by NASA through the Sagan Fellowship Program executed by the NASA Exoplanet Science Institute (R.D.H.)

We acknowledge the support by INAF/Frontiera through the ``Progetti Premiali'' funding scheme of the Italian Ministry of Education, University, and Research.

This publication was made possible through the support of a grant from the John Templeton Foundation. The opinions expressed in this publication are those of the authors and do not necessarily reflect the views of the John Templeton Foundation.

This work was supported in part by a grant from the Carlsberg Foundation.

This paper includes data collected by the \Kepler\ mission. Funding for the \Kepler\ mission is provided by the NASA Science Mission directorate.

Some of the data presented in this paper were obtained from the Mikulski Archive for Space Telescopes (MAST). STScI is operated by the Association of Universities for Research in Astronomy, Inc., under NASA contract NAS5--26555. Support for MAST for non--\textit{Hubble Space Telescope} data is provided by the NASA Office of Space Science via grant NNX13AC07G and by other grants and contracts.

Some of the data presented herein were obtained at the W. M. Keck Observatory (which is operated as a scientific partnership among Caltech, UC, and NASA). The authors wish to recognize and acknowledge the very significant cultural role and reverence that the summit of Maunakea has always had within the indigenous Hawaiian community. We are most fortunate to have the opportunity to conduct observations from this mountain.

We would like to thank the HIRES observers who carried out the RV observations with Keck.

We are grateful to Howard Isaacson for valuable proofreading and suggestions as well as providing access to HIRES $\log R'_{HK}$ observations.

Based on observations made with the Italian {\it Telescopio Nazionale Galileo} (TNG) operated by the {\it Fundaci\'on Galileo Galilei} (FGG) of the {\it Istituto Nazionale di Astrofisica} (INAF) at the {\it  Observatorio del Roque de los Muchachos} (La Palma, Canary Islands, Spain).

The HARPS-N project has been funded by the Prodex Program of the Swiss Space Office (SSO), the Harvard University Origins of Life Initiative (HUOLI), the Scottish Universities Physics Alliance (SUPA), the University of Geneva, the Smithsonian Astrophysical Observatory (SAO), and the Italian National Astrophysical Institute (INAF), the University of St Andrews, Queen’s University Belfast, and the University of Edinburgh.

This work has made use of data from the European Space Agency (ESA) mission \textit{Gaia} (\url{https://www.cosmos.esa.int/gaia}), processed by the \textit{Gaia} Data Processing and Analysis Consortium (DPAC; \url{https://www.cosmos.esa.int/web/gaia/dpac/consortium}). Funding for the DPAC has been provided by national institutions, in particular the institutions participating in the \textit{Gaia} Multilateral Agreement.

Facilities: \facility{Kepler, FLWO:1.5 m (TRES), TNG: (HARPS-N), Keck:I (HIRES), \textit{Gaia}, NASA Exoplanet Archive, ADS, MAST}

Software: \texttt{MultiNest} \citep{ferozetal2009,ferozetal2013}, \texttt{Exo-Transmit} \citep{kemptonetal2017}, \texttt{Pandexo} \citep{batalhaetal2017}, \texttt{JET} (C. D. Fortenbach \& C. D. Dressing 2019, in preparation), \texttt{isochrones} \citep{morton2015a}, \texttt{lightkurve} \citep{barentsenetal2019}, \texttt{BATMAN} \citep{kreidberg2015}

\bibliographystyle{apj}
\bibliography{refs}



\tabletypesize{\small} 
\clearpage 
\fontsize{9}{11}\selectfont
\LongTables
\begin{small}
\begin{deluxetable}{cccccccc}
\tablecolumns{8} 
\tablewidth{0pt}
\tablecaption{RV observations and activity indicators, determined from the DRS.
\label{table:rv_table}} 
\tablehead{ 
BJD & RV & RV error & FWHM & BIS & $\log_{10}(R'_{HK})$ & $\log_{10}(R'_{HK})$ error & Instrument\\
& (m s$^{-1}$) & (m s$^{-1}$) & (km s$^{-1}$) & (km s$^{-1}$) & & &}

\startdata 
$2455402.854339$ & $-8.78$ & $1.32$ & - & - & $-5.078$ & - & HIRES \\
$2455414.971547$ & $-1.24$ & $1.33$ & - & - & $-5.003$ & - & HIRES \\
$2455486.859621$ & $-0.83$ & $1.50$ & - & - & $-4.985$ & - & HIRES \\
$2455544.719659$ & $-5.95$ & $2.14$ & - & - & $-4.971$ & - & HIRES \\
$2455760.087400$ & $ 4.58$ & $1.39$ & - & - & $-4.982$ & - & HIRES \\
$2455796.934228$ & $-3.56$ & $1.25$ & - & - & $-4.974$ & - & HIRES \\
$2455797.920566$ & $-1.33$ & $1.22$ & - & - & $-4.972$ & - & HIRES \\
$2455799.056115$ & $-5.98$ & $1.48$ & - & - & $-5.005$ & - & HIRES \\
$2456114.931149$ & $-6.02$ & $1.27$ & - & - & $-4.944$ & - & HIRES \\
$2456133.896429$ & $-2.53$ & $1.31$ & - & - & $-4.933$ & - & HIRES \\
$2456147.919325$ & $ 0.44$ & $1.31$ & - & - & $-4.967$ & - & HIRES \\
$2456163.912379$ & $ 6.63$ & $1.31$ & - & - & $-4.922$ & - & HIRES \\
$2456164.801818$ & $-0.23$ & $1.30$ & - & - & $-4.924$ & - & HIRES \\
$2456166.047782$ & $ 1.05$ & $1.37$ & - & - & $-4.927$ & - & HIRES \\
$2456166.759374$ & $-2.55$ & $1.44$ & - & - & $-4.913$ & - & HIRES \\
$2456167.990464$ & $-3.55$ & $1.51$ & - & - & $-4.912$ & - & HIRES \\
$2456451.100822$ & $ 2.47$ & $1.48$ & - & - & $-4.953$ & - & HIRES \\
$2456483.086583$ & $ 5.47$ & $1.94$ & - & - & $-4.943$ & - & HIRES \\
$2456486.833662$ & $ 0.39$ & $1.31$ & - & - & $-4.938$ & - & HIRES \\
$2456488.822611$ & $ 7.76$ & $1.19$ & - & - & $-4.926$ & - & HIRES \\
$2456494.987645$ & $ 5.79$ & $1.29$ & - & - & $-4.926$ & - & HIRES \\
$2456506.780605$ & $ 1.27$ & $1.21$ & - & - & $-4.931$ & - & HIRES \\
$2456507.968496$ & $-0.32$ & $1.23$ & - & - & $-4.947$ & - & HIRES \\
$2456532.877110$ & $ 2.73$ & $1.14$ & - & - & $-4.940$ & - & HIRES \\
$2456830.887055$ & $-0.74$ & $1.39$ & - & - & $-4.962$ & - & HIRES \\
$2456850.049952$ & $ 2.59$ & $1.26$ & - & - & $-4.954$ & - & HIRES \\
$2456828.616553$ & $-37327.64$ & $6.07$ & $6.63443$ & $-0.02220$ & $-4.9522$ & $0.0699$ & HARPS-N \\
$2456828.651774$ & $-37320.76$ & $1.50$ & $6.66839$ & $-0.03107$ & $-4.9819$ & $0.0106$ & HARPS-N \\
$2456829.664594$ & $-37319.24$ & $1.56$ & $6.66376$ & $-0.03206$ & $-4.9788$ & $0.0111$ & HARPS-N \\
$2456830.665375$ & $-37319.62$ & $1.83$ & $6.67197$ & $-0.03353$ & $-4.9882$ & $0.0146$ & HARPS-N \\
$2456831.690035$ & $-37319.77$ & $1.70$ & $6.66691$ & $-0.03495$ & $-4.9634$ & $0.0128$ & HARPS-N \\
$2456832.615999$ & $-37314.68$ & $1.80$ & $6.66854$ & $-0.03182$ & $-4.9975$ & $0.0150$ & HARPS-N \\
$2456833.672301$ & $-37322.65$ & $2.12$ & $6.67240$ & $-0.03957$ & $-4.9553$ & $0.0182$ & HARPS-N \\
$2456834.581042$ & $-37315.83$ & $2.01$ & $6.66876$ & $-0.03718$ & $-4.9764$ & $0.0169$ & HARPS-N \\
$2456834.677908$ & $-37321.70$ & $1.98$ & $6.66877$ & $-0.03961$ & $-4.9768$ & $0.0169$ & HARPS-N \\
$2456835.587887$ & $-37318.54$ & $2.07$ & $6.65839$ & $-0.02603$ & $-5.0090$ & $0.0197$ & HARPS-N \\
$2456845.576470$ & $-37322.26$ & $1.44$ & $6.66668$ & $-0.03079$ & $-4.9739$ & $0.0097$ & HARPS-N \\
$2456846.662015$ & $-37327.96$ & $2.10$ & $6.65639$ & $-0.02891$ & $-4.9885$ & $0.0184$ & HARPS-N \\
$2456847.656794$ & $-37321.88$ & $2.67$ & $6.65969$ & $-0.03820$ & $-4.9938$ & $0.0273$ & HARPS-N \\
$2456848.652903$ & $-37327.73$ & $1.68$ & $6.66082$ & $-0.03667$ & $-4.9900$ & $0.0131$ & HARPS-N \\
$2456849.657878$ & $-37326.37$ & $2.17$ & $6.66712$ & $-0.02913$ & $-4.9558$ & $0.0182$ & HARPS-N \\
$2456850.660745$ & $-37324.98$ & $2.17$ & $6.66481$ & $-0.03397$ & $-4.9575$ & $0.0182$ & HARPS-N \\
$2456851.654237$ & $-37323.89$ & $1.66$ & $6.66590$ & $-0.03297$ & $-4.9625$ & $0.0121$ & HARPS-N \\
$2456852.655703$ & $-37316.98$ & $2.55$ & $6.67237$ & $-0.03670$ & $-4.9994$ & $0.0260$ & HARPS-N \\
$2456853.657053$ & $-37318.56$ & $1.62$ & $6.67355$ & $-0.03209$ & $-4.9818$ & $0.0122$ & HARPS-N \\
$2456865.684262$ & $-37320.32$ & $1.67$ & $6.66629$ & $-0.02961$ & $-4.9818$ & $0.0130$ & HARPS-N \\
$2456866.681774$ & $-37323.10$ & $3.48$ & $6.66279$ & $-0.04100$ & $-4.9629$ & $0.0388$ & HARPS-N \\
$2456883.639193$ & $-37324.45$ & $1.89$ & $6.65864$ & $-0.03904$ & $-5.0004$ & $0.0164$ & HARPS-N \\
$2456884.647365$ & $-37324.61$ & $1.84$ & $6.66539$ & $-0.02953$ & $-4.9982$ & $0.0166$ & HARPS-N \\
$2456885.644031$ & $-37322.71$ & $1.86$ & $6.66503$ & $-0.03266$ & $-4.9753$ & $0.0159$ & HARPS-N \\
$2456886.642561$ & $-37346.88$ & $11.81$ & $6.66764$ & $-0.08202$ & $-5.0609$ & $0.2193$ & HARPS-N \\
$2456887.651622$ & $-37322.33$ & $1.83$ & $6.67033$ & $-0.03390$ & $-4.9883$ & $0.0152$ & HARPS-N \\
$2456888.580937$ & $-37321.19$ & $2.74$ & $6.65114$ & $-0.02926$ & $-4.9577$ & $0.0266$ & HARPS-N \\
$2456889.585275$ & $-37324.05$ & $2.15$ & $6.65998$ & $-0.03761$ & $-4.9716$ & $0.0189$ & HARPS-N \\
$2456903.541993$ & $-37318.65$ & $1.41$ & $6.66569$ & $-0.03568$ & $-4.9851$ & $0.0095$ & HARPS-N \\
$2456919.514886$ & $-37322.96$ & $2.44$ & $6.66569$ & $-0.03738$ & $-4.9735$ & $0.0224$ & HARPS-N \\
$2456922.547287$ & $-37323.20$ & $1.67$ & $6.65621$ & $-0.04098$ & $-5.0131$ & $0.0146$ & HARPS-N \\
$2456923.501548$ & $-37320.74$ & $1.62$ & $6.66346$ & $-0.03370$ & $-5.0026$ & $0.0127$ & HARPS-N \\
$2456924.510113$ & $-37318.66$ & $2.43$ & $6.65831$ & $-0.03765$ & $-4.9763$ & $0.0237$ & HARPS-N \\
$2456936.514073$ & $-37326.01$ & $1.72$ & $6.65525$ & $-0.03770$ & $-4.9864$ & $0.0143$ & HARPS-N \\
$2456939.418861$ & $-37323.19$ & $1.36$ & $6.65561$ & $-0.03499$ & $-4.9986$ & $0.0093$ & HARPS-N \\
$2456969.402685$ & $-37323.75$ & $3.10$ & $6.65044$ & $-0.03308$ & $-5.0094$ & $0.0377$ & HARPS-N \\
$2457106.734166$ & $-37320.49$ & $2.77$ & $6.66677$ & $-0.03337$ & $-4.9801$ & $0.0289$ & HARPS-N \\
$2457116.717298$ & $-37324.15$ & $1.51$ & $6.64918$ & $-0.03136$ & $-5.0282$ & $0.0120$ & HARPS-N \\
$2457118.706394$ & $-37327.91$ & $1.87$ & $6.64875$ & $-0.03660$ & $-5.0338$ & $0.0181$ & HARPS-N \\
$2457121.726137$ & $-37324.71$ & $1.61$ & $6.64485$ & $-0.03692$ & $-5.0174$ & $0.0141$ & HARPS-N \\
$2457153.685174$ & $-37323.84$ & $2.87$ & $6.64814$ & $-0.04127$ & $-5.0063$ & $0.0321$ & HARPS-N \\
$2457156.714776$ & $-37324.15$ & $12.94$ & $6.66453$ & $-0.05998$ & $-4.9161$ & $0.1624$ & HARPS-N \\
$2457159.642662$ & $-37323.22$ & $2.14$ & $6.65282$ & $-0.03562$ & $-4.9989$ & $0.0202$ & HARPS-N \\
$2457160.638323$ & $-37320.99$ & $2.03$ & $6.65589$ & $-0.04059$ & $-5.0130$ & $0.0185$ & HARPS-N \\
$2457161.626357$ & $-37324.63$ & $1.70$ & $6.65201$ & $-0.04164$ & $-5.0095$ & $0.0140$ & HARPS-N \\
$2457180.658376$ & $-37322.50$ & $1.60$ & $6.64953$ & $-0.03798$ & $-5.0160$ & $0.0127$ & HARPS-N \\
$2457181.686408$ & $-37322.02$ & $1.75$ & $6.65033$ & $-0.03871$ & $-5.0016$ & $0.0146$ & HARPS-N \\
$2457182.670828$ & $-37322.60$ & $1.58$ & $6.64797$ & $-0.03663$ & $-5.0014$ & $0.0122$ & HARPS-N \\
$2457183.652886$ & $-37321.34$ & $1.81$ & $6.64944$ & $-0.03550$ & $-4.9890$ & $0.0145$ & HARPS-N \\
$2457184.643705$ & $-37324.94$ & $2.35$ & $6.65649$ & $-0.03450$ & $-5.0265$ & $0.0237$ & HARPS-N \\
$2457185.662466$ & $-37324.52$ & $1.51$ & $6.65258$ & $-0.03701$ & $-5.0096$ & $0.0115$ & HARPS-N \\
$2457186.662672$ & $-37325.81$ & $1.52$ & $6.65510$ & $-0.03815$ & $-5.0237$ & $0.0118$ & HARPS-N \\
$2457188.679310$ & $-37328.26$ & $1.58$ & $6.65099$ & $-0.03864$ & $-5.0010$ & $0.0123$ & HARPS-N \\
$2457189.672084$ & $-37322.40$ & $2.63$ & $6.64860$ & $-0.03852$ & $-5.0649$ & $0.0316$ & HARPS-N \\
$2457190.685669$ & $-37325.59$ & $1.61$ & $6.64984$ & $-0.03343$ & $-5.0211$ & $0.0135$ & HARPS-N \\
$2457191.685746$ & $-37323.07$ & $1.57$ & $6.65434$ & $-0.03606$ & $-5.0243$ & $0.0128$ & HARPS-N \\
$2457192.684342$ & $-37324.65$ & $1.67$ & $6.65251$ & $-0.03217$ & $-5.0374$ & $0.0147$ & HARPS-N \\
$2457193.684869$ & $-37323.90$ & $1.44$ & $6.65760$ & $-0.03912$ & $-5.0097$ & $0.0107$ & HARPS-N \\
$2457195.594752$ & $-37318.06$ & $1.65$ & $6.65239$ & $-0.03455$ & $-4.9930$ & $0.0132$ & HARPS-N \\
$2457221.626801$ & $-37324.84$ & $1.37$ & $6.64834$ & $-0.04333$ & $-5.0237$ & $0.0101$ & HARPS-N \\
$2457222.569536$ & $-37324.99$ & $1.78$ & $6.65155$ & $-0.04076$ & $-4.9982$ & $0.0150$ & HARPS-N \\
$2457223.579194$ & $-37322.72$ & $3.17$ & $6.63767$ & $-0.04077$ & $-5.0049$ & $0.0375$ & HARPS-N \\
$2457225.522395$ & $-37319.94$ & $3.63$ & $6.64970$ & $-0.03534$ & $-5.0279$ & $0.0478$ & HARPS-N \\
$2457226.582966$ & $-37317.08$ & $3.17$ & $6.66790$ & $-0.02615$ & $-5.0750$ & $0.0424$ & HARPS-N \\
$2457226.606265$ & $-37321.98$ & $2.58$ & $6.65109$ & $-0.03404$ & $-5.0105$ & $0.0277$ & HARPS-N \\
$2457227.627333$ & $-37316.52$ & $1.67$ & $6.66113$ & $-0.03348$ & $-5.0084$ & $0.0136$ & HARPS-N \\
$2457228.630703$ & $-37320.68$ & $2.72$ & $6.66677$ & $-0.02712$ & $-4.9272$ & $0.0251$ & HARPS-N \\
$2457229.584812$ & $-37316.53$ & $2.51$ & $6.65780$ & $-0.02953$ & $-5.0054$ & $0.0266$ & HARPS-N \\
$2457230.528273$ & $-37320.18$ & $3.34$ & $6.66154$ & $-0.02550$ & $-4.9889$ & $0.0385$ & HARPS-N \\
$2457254.631919$ & $-37325.16$ & $2.41$ & $6.66627$ & $-0.03812$ & $-5.0093$ & $0.0260$ & HARPS-N \\
$2457256.398580$ & $-37314.12$ & $2.36$ & $6.64870$ & $-0.04351$ & $-4.9894$ & $0.0239$ & HARPS-N \\
$2457257.413714$ & $-37315.46$ & $1.88$ & $6.65391$ & $-0.03681$ & $-4.9651$ & $0.0149$ & HARPS-N \\
$2457267.435503$ & $-37324.84$ & $2.04$ & $6.64591$ & $-0.03502$ & $-5.0136$ & $0.0193$ & HARPS-N \\
$2457268.492540$ & $-37322.86$ & $2.21$ & $6.64810$ & $-0.04496$ & $-5.0215$ & $0.0224$ & HARPS-N \\
$2457269.418733$ & $-37320.32$ & $1.69$ & $6.65385$ & $-0.03696$ & $-5.0211$ & $0.0145$ & HARPS-N \\
$2457270.407599$ & $-37321.14$ & $1.36$ & $6.65014$ & $-0.03809$ & $-5.0301$ & $0.0099$ & HARPS-N \\
$2457271.408119$ & $-37325.01$ & $1.44$ & $6.65177$ & $-0.03459$ & $-5.0336$ & $0.0110$ & HARPS-N \\
$2457273.426969$ & $-37326.84$ & $1.46$ & $6.64860$ & $-0.04486$ & $-5.0183$ & $0.0109$ & HARPS-N \\
$2457301.384627$ & $-37319.51$ & $1.47$ & $6.65573$ & $-0.03632$ & $-5.0121$ & $0.0111$ & HARPS-N \\
$2457302.383904$ & $-37318.58$ & $1.79$ & $6.65019$ & $-0.03970$ & $-5.0308$ & $0.0159$ & HARPS-N \\
$2457321.426080$ & $-37318.14$ & $2.40$ & $6.64231$ & $-0.03293$ & $-5.0135$ & $0.0243$ & HARPS-N \\
$2457330.417736$ & $-37318.77$ & $1.89$ & $6.65257$ & $-0.03583$ & $-5.0053$ & $0.0167$ & HARPS-N \\
$2457334.397358$ & $-37321.22$ & $1.64$ & $6.65030$ & $-0.04224$ & $-5.0085$ & $0.0131$ & HARPS-N
\enddata
\end{deluxetable}
\clearpage 
\end{small}


\clearpage

\end{document}